\documentclass[aps,prd,twocolumn,groupedaddress, showkeys, showpacs]{revtex4}
\usepackage{graphicx,subfigure,enumerate,pstricks,latexsym,longtable,amssymb}
\usepackage[english]{babel}

\newcommand{\Schw}{Schwarzschild}

\newcommand{\beq}{\begin{equation}}
\newcommand{\eeq}{\end{equation}}
\newcommand{\bea}{\begin{eqnarray}}
\newcommand{\eea}{\end{eqnarray}}

 \newcommand{\cl}{{\cal{L}}} \newcommand{\cb}{{\cal{B}}}

\def\lin{ --- }

\def\d{\dif}

\def\EE{{\cal E}}
\def\LL{{\cal L}}
\def\BB{{\cal B}}
\def\QQ{{\cal Q}}

\def\RS{\Sigma}

\def\Omk{\Omega_{ K}}

\providecommand{\dif}{\mathrm{d}} \def\d{\dif}

\def\aprx{\sim}

\begin{document}

\title{Possible signature of magnetic fields related to quasi-periodic oscillations observed in microquasars}

\author{Martin Kolo\v{s}}
\email{martin.kolos@fpf.slu.cz}
\author{Arman Tursunov}
\email{arman.tursunov@fpf.slu.cz}
\author{Zden\v{e}k Stuchl{\'i}k}
\email{zdenek.stuchlik@fpf.slu.cz}

\affiliation{Institute of Physics and Research Centre of Theoretical Physics and Astrophysics, Faculty of Philosophy and Science, Silesian University in Opava, \\
Bezru{\v c}ovo n{\'a}m.13, CZ-74601 Opava, Czech Republic}
\begin{abstract}
The study of quasi-periodic oscillations (QPOs) of X-ray flux observed in the stellar-mass black hole binaries can provide a powerful
tool for testing of the phenomena occurring in strong gravity regime. Magnetized versions of the standard geodesic models of QPOs can explain the observationally fixed data from the three microquasars. We perform a successful fitting of the HF QPOs observed in three microquasars, GRS 1915+105, XTE 1550-564 and GRO 1655-40, containing black holes, for magnetised versions of both epicyclic resonance and relativistic precession models and discuss the corresponding constraints of parameters of the model, which are the mass and spin of the black hole and the parameter related to the external magnetic field. 
The estimated magnetic field intensity strongly depends on the type of objects giving the observed HF QPOs. It can be as small as $10^{-5}$~G if electron oscillatory motion is relevant, but it can be by many orders higher for protons or ions ($0.02-1$~G), or even higher for charged dust or such exotic objects as lighting balls, etc. On the other hand, if we know, for some reasons, magnetic field intensity, our model implies strong limit on character of the oscillating matter, namely its specific charge.
\end{abstract}

\pacs{04.70.Bw, 04.25.-g, 04.70.-s, 97.60.Lf \hfill
}

\maketitle

\section{Introduction} \label{intro}

Microquasars are binary systems composed of a black hole and a companion (donor) star; matter floating from the companion star onto the black hole forms an accretion disk and relativistic jets - bipolar outflow of matter along the black hole - accretion disk rotation axis. Due to friction, matter of the accretion disk becomes to be hot and emits electromagnetic radiation, also in X-rays in vicinity of the black hole horizon.

Applying the methods of spectroscopy (frequency distribution of photons) and timing (photon number time dependence) for particular microquasars, one can extract a useful information regarding the range of parameters of the system \citep{Rem-McCli:2006:ARAA:}. In this connection, the binary systems containing black holes, being compared to neutron star systems, seem to be promising due to the reason that any astrophysical black hole is thought to be a Kerr black hole (corresponding to the unique solution of general relativity in 4D for uncharged black holes which does not violate the no hair theorem and the weak cosmic censorship conjecture) that is determined by only two parameters: the black hole mass $M$ and the dimensionless spin $|a| \leq 1$. 

One of the promising tools to probe the phenomena occurring in the field of the black hole candidates is the study of the quasi-periodic oscillations (QPOs) of the X-ray power density observed in microquasars. The current technical possibilities to measure the frequencies of QPOs with high precision allow us to get useful knowledge about the central object and its background. According to the observed frequencies of QPOs, which cover the range from few mHz up to $0.5$kHz, different types of QPOs were distinguished. Mainly, these are the high frequency (HF) and low frequency (LF) QPOs with frequencies up to 500~Hz and up to 30~Hz, respectively. The HF QPOs in black hole microquasars are usually detected with the twin peaks which have frequency ratio close to $3:2$. This is the case of the Galactic black hole microquasars GRS 1915+105, XTE 1550-564 and GRO J1655-40, that we consider here. 

After the first detection of QPOs, there were many attempts to fit the observed QPOs, and different models have been proposed, such as the hot-spot models, disko-seismic models, warped disk model and many versions of resonance models, developed in the framework of general relativity or alternative theories of gravity. It is particularly interesting that the characteristic frequencies of HF QPOs are close to the values of the frequencies of test particle, geodesic epicyclic oscillations in the regions near the innermost stable circular orbit (ISCO) which makes it reasonable to construct the model involving the frequencies of oscillations associated with the orbital motion around Kerr black holes. However, until now, the exact physical mechanism of the generation of HF QPOs is not known, since none of the models can fit the observational data from different sources. Surprisingly, the situation changes considerably if one assumes the oscillating matter to be electrically charged, and takes into account the influence of external magnetic fields in the curved black hole background. This is the main aim of the present paper, such assumptions are relevant and astrophysically motivated due to several reasons described below.

Magnetic fields have been detected and measured in nearly all celestial objects. The existence of accretion disk around black hole which usually contains highly conducting plasma can lead to the appearance of the regular electromagnetic fields in the vicinity of a black hole. Jets from stellar-mass compact objects such as the black hole microquasars are believed to be generated due to the interaction of charged matter with the gravitational field of a black hole combined with surrounding electromagnetic fields, rather than the radiation pressure. Even thought the total charge of an accretion disk is believed to be zero, the presence of plasma implies the existence of local charges in small limited regions along an accretion disk which may be influenced by the external magnetic fields \citep{Cre-etal:2013:ApJS:}. Moreover, the collimation of jets at large distances supports the idea of the existence of large-scale magnetic fields in the interstellar medium. 

This implies that the role of magnetic fields in the vicinity of astrophysical black holes cannot be neglected. Typically, the Galactic large-scale magnetic fields are weak, as can be estimated from the intensity of total synchrotron radiation producing the X-rays, or by methods including the Faraday rotation of the plane of polarization of a starlight in optical range propagating in a magnetized plasma, or by the Zeeman effect in the vicinity of stars. The average equipartition strength of the magnetic field in spiral galaxies is estimated to be of order $\sim 10^{-5}$~G. The strength of the magnetic field increases closer to the center of the Galaxy, achieving from tens up to hundred Gauss around supermassive black holes \citep{Eat-etal:2013:NATUR:}. The recent studies of the structure of magnetic fields in galaxies \citep{Bec-Wie:2013:book}, clearly demonstrate the formation of the spiral patterns in almost all considered galaxies, including Milky Way. 
Therefore, black holes can be also immersed into the external, large scale magnetic fields that can have globally complicated structure, but at large distances from its source, in a finite element of space, its character can be considered as locally uniform.

We concentrate our attention on the particular and simplified case of weak magnetic fields that are considered to be asymptotically uniform at spatial infinity, described by the well known Wald solution for weakly magnetized black holes \citep{Wald:1974:PHYSR4:}. The condition of the weakness of magnetic field implies the fact that its stress-energy tensor does not violate the geometry of the black hole spacetime. In this sense the test field approximation can be applied when the strength of the magnetic field in the vicinity of a black hole with mass $M$ fulfills the following condition (see, e.g., \cite{Fro-Sho:2010:PHYSR4:})
\beq \label{BBB}
B << B_{\rm G}=\frac{c^4}{G^{3/2} M_{\odot}} \left(\frac{{M}_{\odot}}{M}\right)\aprx10^{19}{ \frac{{M}_{\odot}} {M} }\,{\rm Gauss}\, .
\eeq
Obviously, for astrophysical black holes the weak field condition (\ref{BBB}) is perfectly satisfied. However, as we will see in the following sections, for the motion of charged test particles the magnetic field plays a crucial role due to the large factor of the specific charge $q/m$ entering the equations.

In the present paper, we consider the motion of charged test particles around a Kerr black hole immersed into an external asymptotically homogeneous magnetic field having the field lines aligned with the black hole rotation axis. We look especially for the existence and properties of the harmonic or quasi-harmonic oscillations of charged particles in the magnetized black hole backgrounds. The quasi-harmonic oscillations around a stable equilibrium location and the frequencies of these oscillations are then compared with the frequencies of the HF and LF QPOs observed in the three particular microquasars GRS 1915+105, XTE 1550-564, and GRO 1655-40 \citep{McC-etal:2011:CLAQG:,Tor-etal:2005:ASTRA:}. 
 We demonstrate that the effect of the external magnetic field, which is usually neglected in most of the QPO models, allows us to fit the observed QPO frequencies for different sources in a single model. Assuming the detected X-rays in QPOs are generated due to the synchrotron radiation of electrons, we estimate the magnetic field strength necessary and sufficient for the fitting of the QPOs frequencies. We show that the magnetic field strength in this case is of the order which is comparable to the magnitude of the Galactic magnetic field. In principle, the oscillating matter could be constituted from heavier particles, protons, ions, charged dust, or some exotic objects similar to ball lighting, proposed in \cite{Stu-Kot-Tor:2013:ASTRA:}; for such possible sources, the magnetic field should be significantly stronger than in the electron case, in dependence on specific charge.

Influence of the magnetic field on the charged test particle motion around black holes has been already studied in literature \citep{Fro-Sho:2010:PHYSR4:, Kon-Liu:2012:PRD:,Kov-etal:2010:CLAQG:,Kop-Kar:2014:APJ:}. Charged particle motion in an uniform magnetic field, and related high frequency oscillations, have been studied for \Schw{} black holes in \cite{Kol-Stu-Tur:2015:CLAQG:}, locally measured angular frequencies for the magnetized Kerr black hole with the dynamics of charged particles were studied in \cite{Tur-Stu-Kol:2016:PRD:}. A complementary simple model of creation of charged jets has been introduced in \cite{Stu-Kol:2016:EPJC:}. 
For charged particle motion in the Reissner-Nordstr{\"o}m and Kerr-Newman backgrounds see \citep{Pug-Que-Ruf:2011:PHYSR4:,Pug-Que-Ruf:2017:EPJC:,Bic-Stu-Bal:1989:BAC:,Bal-Bic-Stu:1989:BAC:,Stu-Bic-Bal:1999:GRG:}.
The effect of the dipole magnetic field on the oscillatory charged particle motion has been studied in \citep{Bak-etal:2010:CLAQG:,Bak-etal:2012:CLAQG:}. The dipole magnetic field configuration is assumed to be much more relevant for neutron star X-ray binaries, while for black hole X-ray binaries we can use the uniform configuration of the magnetic field.
The influence of the magnetic field on the modes of the diskoseismic oscillations of an accretion disk orbiting a black hole, and its relation to observed HF QPOs, has been studies in \cite{Fu-Lai:2009:ApJ:}.

Throughout the present paper, we use the spacelike signature $(-,+,+,+)$, and the system of geometric units in which $G = 1 = c$. However, for expressions having astrophysical relevance we use the physical constants explicitly. Greek indices are taken to run from 0 to 3.

\section{Charged particle motion in the field of magnetized Kerr black hole} \label{WMKBH}

We start from the general description of the motion of a charged test particle in the field of a rotating black hole in the presence of the external magnetic field. The geometry of the rotating black hole is given by the Kerr metric 
\beq 
\d s^2 = g_{tt}\d{t}^2 + 2g_{t\phi}\d{t}\d\phi + g_{\phi\phi}\d\phi^2 + g_{rr}\d{r}^2 + g_{\theta\theta}\d\theta^2 , \label{KerrMetric}
\eeq
with the nonzero components of the metric tensor $g_{\mu\nu}$ taking in the standard Boyer-Lindquist coordinates the form 
\bea 
g_{tt} = - \left( 1- \frac{2Mr}{\RS} \right), \quad
g_{t\phi} = - \frac{2Mra \sin^2\theta}{\RS} , \nonumber\\
g_{\phi\phi} = \left( r^2 +a^2 + \frac{2Mra^2}{\RS} \sin^2\theta \right) \sin^2\theta, \nonumber \\
g_{r r} = \frac{\RS}{\Delta}, \quad g_{\theta\theta} = \RS,
\label{MetricCoef} 
\eea
where 
\beq 
\RS = r^2 + a^2 \cos^2\theta, \quad \Delta = r^2 - 2Mr + a^2. 
\eeq
Here, $M$ is the gravitational mass of the black hole and $a$ is its spin parameter. The physical singularity is located at the ring with $r=0, \quad \theta = \pi/2$. The outer horizon is located at 
\beq
r_{+} = M + (M^2-a^2)^{1/2}.
\eeq

\subsection{Uniform external magnetic field}

The external axially symmetric and asymptotically homogeneous magnetic field implies the electromagnetic field potential which is solution of the vacuum Maxwell equations taking in general the form \citep{Wald:1974:PHYSR4:}
\beq 
A^{\alpha} = \frac{B}{2} \left(\xi_{(\phi)}^{\alpha} + 2 a \xi_{(t)}^{\alpha}\right) - \frac{Q}{2M} \xi_{(t)}^{\alpha}. \label{VecPotAcr}
\eeq
Here $Q$ refers to the induced electric potential which exists due to the rotation of the black hole giving the contribution to the Faraday induction. This causes the process of a selective accretion of charged particles into the rotating black hole. However the stage of the selective accretion ends shortly in the most of the astrophysical scenarios when the black hole gets the induced charge $Q=Q_{W}\equiv 2 a M B$. The charge $Q_{W}$ is called the Wald charge. Finally, the four-vector potential of an electromagnetic field  after the process of selective accretion completed takes form $A^{\alpha} = \frac{1}{2} B \xi_{(\phi)}^{\alpha}$, where $B$ denotes the intensity of the asymptotically uniform magnetic fiels; $\xi_{(\phi)}, \xi_{(t)}$ denote the axial and time Killing vector fields.
Hereafter, we will consider two limiting cases with the induced charge $Q$ of a black hole: $Q=0$ and $Q=Q_{\rm W}$. So, the four vector potential of electromagnetic field for these two cases takes the form
\beq  A_0^{\alpha} = \frac{B}{2} \left(\xi_{(\phi)}^{\alpha} + 2 a \xi_{(t)}^{\alpha}\right), \quad A_{\rm W}^{\alpha} = \frac{B}{2} \xi_{(\phi)}^{\alpha}.  \label{VecPotShort}
\eeq
One can neglect the gravitational effect of the induced charge, if the following condition is satisfied \citep{Tur-Stu-Kol:2016:PRD:} 
\beq 
Q << Q_{\rm G} = 2 G^{1/2} M \approx 10^{31} \frac{M}{10 M_{\odot}} ~\mbox{statC}. \label{QG} 
\eeq 
This condition holds well in our approach since the maximal induced charge of a magnetized rotating black hole, namely $Q_{\rm W}$, is 12 orders of magnitude less than $Q_{\rm G}$ for the solar mass black holes.

The electromagnetic field tensor reads $F_{\mu\nu} = A_{\nu,\mu} - A_{\mu,\nu}$. For the potential (\ref{VecPotAcr}), and in the equatorial plane $\theta=\pi/2$, one can write the nonvanishing independent components of $F_{\mu\nu}$ as
\beq F_{r \phi} = \frac{B (r^3 \pm a^2)}{r^2}, \quad F_{r t} = \mp \frac{a B}{r^2},
\eeq
where the upper and lower signs correspond to the cases with $Q=0$ and $Q=Q_W$, respectively. The expressions for the Maxwell tensor $F_{\mu\nu}$ in the equatorial plane will be useful in the next section for the calculation of the frequencies of the charged particle epicyclic oscillations. In general, the antisymmetric tensor $F_{\mu\nu}$ has 4 nonzero components in the magnetized rotating black hole case.

In order to describe the dynamics of the charged particle motion, we introduce the following dimensionless parameters
\beq 
 \quad \BB = \frac{q B M}{2m},  \qquad \QQ = \frac{q Q}{2 m M}.  \label{BBdef} 
\eeq
Putting hereafter for the shortness $M=1$, the Wald charge corresponds to $\QQ_W = 2 a \BB$. The parameter $\BB$ we will be called the magnetic parameter \citep{Tur-Stu-Kol:2016:PRD:}. The estimation of the magnitude of the magnetic parameter shows that the effect of even weak magnetic field cannot be neglected for the charged particle motion due to the large value of the specific charge $q/m$ of the test particle. For example, for electrons in the vicinity of a stellar mass black hole $M\approx{10 {M}_{\odot}}$ with the magnetic parameter $\BB\approx{1}$, the magnetic field strength corresponds to the value of $B\approx0.002$~G; for protons in the same conditions, the strength of magnetic the field $B$ has to be $1836$ times stronger. More detailed analysis of the magnetic parameter in astrophysical scenarios related to the generation of QPOs observed in the black hole microquasars will be given in Sec.~\ref{observations}.

\subsection{Circular obits of charged particles}

The motion of a charged particle with charge $q$ and mass $m$ in curved spacetime in the presence of electromagnetic field can be treated by the Lorentz equation
\beq 
 \frac{\d u^\mu}{\d \tau} + \Gamma_{\alpha\beta}^{\mu} u^{\alpha} u^{\beta} = \frac{q}{m} g^{\mu\rho} F_{\rho\sigma} u^{\sigma}, \label{geomageq} 
\eeq
where $u^{\mu} = dx^{\mu} / d\tau$ is the four-velocity of the particle. The regular motion of charged particles in magnetized black hole backgrounds is always bounded in radial direction which makes important to study the cyclic motion of the particles. In particular, the circular motion of a particle is possible in the equatorial plane $\theta = \pi/2$, which follows from the properties of the symmetry of the geometry and background electromagnetic field. 

The four velocity of the circular motion has only 2 nonvanishing components, $u^{\mu} = \{u^{t},0,0,u^{\phi}\}$. This implies that for the equatorial circular motion, the radial component of Eq. (\ref{geomageq}) can be written in the form
\beq
(a^2 - r^3) (u^{\phi})^2 - 2 \BB (r^3 \pm a^2) u^{\phi} -  2 a (u^{\phi} \mp \BB) u^{t} + (u^{t})^2 = 0, \label{geomageqR}
\eeq
where the upper sign corresponds to the case with $\QQ = Q = 0$, and the lower sign to the case with  $\QQ = \QQ_W$.
The normalization condition $g_{\mu\nu} u^{\mu} u^{\nu} = -1$ gives the second equation for the nonzero components of the four-velocity in the form
\beq 
(a^2 (2 + r) + r^3) (u^{\phi})^2 - 4 a u^{\phi} u^{t} - (r-2) (u^{t})^2 + r = 0, \label{norcon}
\eeq
which is obviously independent of the electromagnetic parameters $\BB$ and $\QQ$. Eq. (\ref{geomageqR}) and (\ref{norcon}) allow us to find two expressions for two unknown quantities $u^{t}$ and $u^{\phi}$. The explicit form of analytical expressions for $u^{t}$ and $u^{\phi}$ cannot be represented in a simple form, however we can easily solve it numerically. Results of the numerical calculations will be demonstrated in the following section. This form of the equations of motion will be useful for the calculation of frequencies of the quasi-harmonic oscillations of charged particles along circular orbits in the field of magnetized black holes. 

One can find the equations of motion in a different way using the Hamiltonian formalism. Due to the symmetries of the Kerr geometry and the axial symmetry of an external magnetic field, the components of four velocity are related to the conservation of the following quantities
\bea
 - \EE &=& \frac{P_t}{m} = g_{tt} u^{t} + g_{t\phi} u^{\phi} + \frac{q}{m} A_t, \\
 \LL &=& \frac{P_{\phi}}{m} = g_{\phi\phi} u^{\phi} + g_{t\phi} u^{t} + \frac{q}{m} A_\phi, \label{enangmom}
\eea
where $\EE = E/m$ and $\LL = L/m$ are identified as the specific energy and specific angular momentum of the charged particle, respectively. This gives us a right to write the Hamiltonian of the charged particle motion in the form \citep{Wald:1984:book:}
\beq
  H_{\rm p} =  \frac{1}{2} g^{\alpha\beta} (P_\alpha - q A_\alpha)(P_\beta - q A_\beta) + \frac{1}{2} \, m^2
  \label{particleHAM},
\eeq
where $P_{\mu}$ is a generalized (canonical) four-momentum related to the four-velocity by the relation
\beq
 P^\mu = m u^\mu + q A^\mu. \label{particleMOM}
\eeq
Equations of motion can be then written, in the form of the Hamilton equations
\beq
\frac{\d x^{\mu}}{\d\zeta} \equiv m u^{\mu} = \frac{\partial H}{\partial P_\mu}, \quad
 \frac{\d P_\mu}{\d \zeta} = - \frac{\partial H}{\partial x^\mu}, \label{Ham_eq}
\eeq
where the affine parameter $\zeta = \tau/m$ is related to the proper time of the particle $\tau$. The dynamics of charged particles in magnetized rotating black hole background was studied in detail in \cite{Tur-Stu-Kol:2016:PRD:}, where we use the Hamiltonian formalism combined with the so called formalism of forces which allows us to find the analytical solutions for the motion of charged particles. Stability of the circular orbits is also discussed in \cite{Tur-Stu-Kol:2016:PRD:} - the harmonic quasi-periodic oscillations are allowed only around the stable circular orbits.

\section{Harmonic oscillations as perturbation of circular orbits} \label{oscillations}

In order to describe the oscillatory motion of charged particle we use the perturbation of the equations of motion around the stable circular orbits. If a charged test particle is slightly displaced from the equilibrium position $x^{\mu}_{0}$ corresponding to a stable circular orbit located at the equatorial plane, the particle will start to oscillate around the stable orbit realizing thus epicyclic motion governed by linear harmonic oscillations. Following \cite{Ali-Gal:1981:GRG:}, one can introduce the deviation vector $\xi^{\mu}(\tau) = x^{\mu}(\tau) - x^{\mu}_{0}(\tau)$. Substituting the deviation vector into the equation (\ref{geomageq}) and using the first order of expansion, we get the equation for $\xi^{\mu}$ in the form
\beq \frac{d^2 \xi^{s}}{dt} + \Omega^2_s \xi^s = 0, \quad s = r, \theta, \eeq
where $\Omega_r,\Omega_\theta$ are the epicyclic frequencies of the particle oscillations near the black hole, measured by a distant static observer. 
The  radial and vertical oscillatory frequencies for the case $Q=Q_W=2 a \BB$ take the explicit form
\bea
\Omega_\theta^2 &=& \frac{\alpha u^\phi (u^\phi + 2 \BB) + \beta u^t (u^\phi + \BB) + 2 a^2 (u^t)^2}{r^5 (u^t)^2} , \label{Omtheta} \\
\Omega_r^2 &=& \frac{\gamma u^\phi (u^\phi + 2 \BB) + \mu u^t (u^\phi + \BB) + \rho (u^t)^2 + \sigma}{r^5 (u^t)^2}, \label{Omradial} 
\eea
where the introduced coefficients read
\bea 
\alpha &=& r^5 + a^2 r^2 (4 + r) + 2 a^4, \nonumber \\ 
\beta &=& -4 a (r^2 + a^2), \nonumber \\
\gamma &=& r^4 (-8 + 3 r) + a^2 r (2 - 10 r + r^2) - 4 a^4, \nonumber \\
\mu &=& 4 a (2 a^2 - r + 3 r^2), \nonumber \\
\rho &=& 2 (-2 a^2 + r - r^2), \nonumber \\
\sigma &=& 4 \BB^2 r ((-2 + r) r^3 - a^2 (1 + 2 r)),
\eea
while $u^\phi$ and $u^{t}$ are the components of the four velocity of the oscillating particle given by the solution of Eq. (\ref{geomageqR}) and (\ref{norcon}).
The frequencies for the case with the zero induced charge will be similar to the Eq. (\ref{Omtheta}) and (\ref{Omradial}). The difference between the frequencies of the oscillations in the presence, $\Omega_{\rm(W)}$, and the absence, $\Omega_{\rm(0)}$, of the Wald charge are given by the following expressions
\bea
\Omega_{\theta {\rm(W)}}^2  - \Omega_{\theta (0)}^2 &=& \frac{8 a^2 \BB}{r^5 (u^t)^2} \left((a^2 + r^2) u^{\phi} - a u^t\right), \label{diffQWandQ0a} \\
\Omega_{r {\rm(W)}}^2  - \Omega_{r (0)}^2 &=& \frac{-2 a \BB}{r^5 (u^t)^2} \left(\mu u^{\phi} + 2 \rho u^t + 8 a \BB r^2 \right). \label{diffQWandQ0b}
\eea
Note that the epicyclic frequencies (\ref{Omtheta}) and (\ref{Omradial}) are given for the oscillations measured from the rest at infinity. Locally measured frequencies of the radial and latitudinal harmonic oscillations were presented in our previous paper \citep{Tur-Stu-Kol:2016:PRD:} in terms of the energy and angular momentum of a charged particle at the circular orbit. In addition to the epicyclic frequencies $\Omega_r$ and $\Omega_{\theta}$, there exist also the Keplerian frequency, $\Omk$, and so called Larmor angular frequency, $\Omega_{L}$, associated with the pure contribution of an external uniform magnetic field, that are given by the relations
\beq
 \Omk \equiv \Omega_{\phi} = \frac{d \phi}{d t} = \frac{u^\phi}{u^{t}}, \qquad \Omega_{L} \equiv \frac{q B}{m u^t} = \frac{2\BB}{u^t}. \label{OmKepLar} 
\eeq
The Larmor angular frequency depends only on the strength of the magnetic parameter $\BB$ and the redshift factor $u^t$. It is fully relevant in large distances from the black hole where influence of the uniform magnetic field becomes to be crucial.

\begin{figure*}
\includegraphics[width=\hsize]{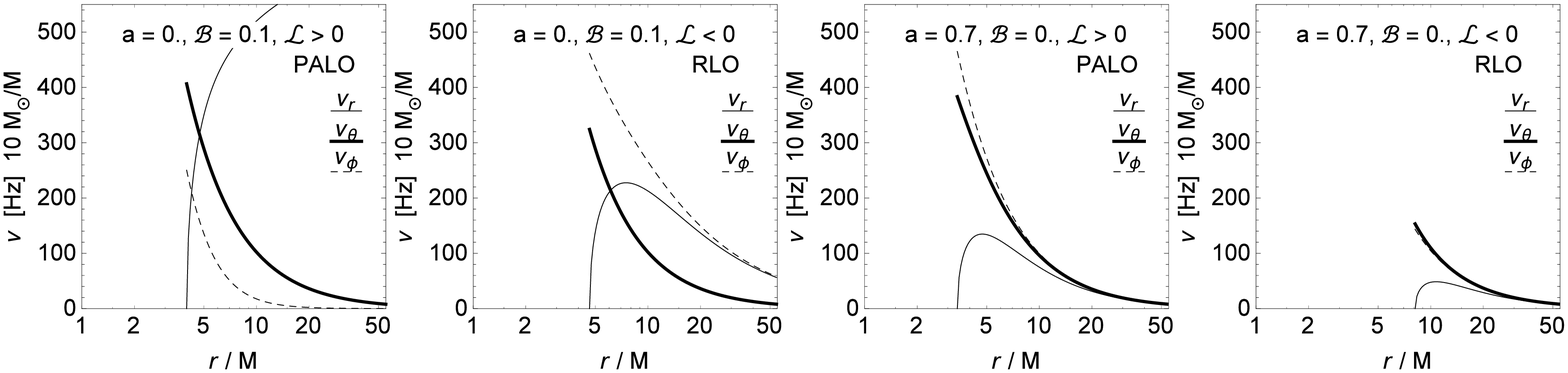}
\includegraphics[width=\hsize]{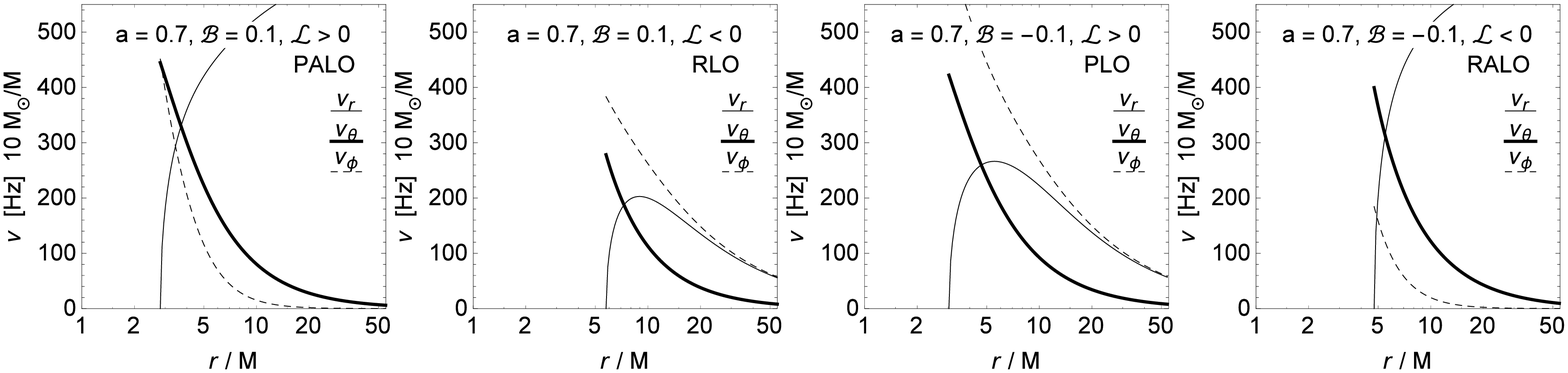}
\includegraphics[width=\hsize]{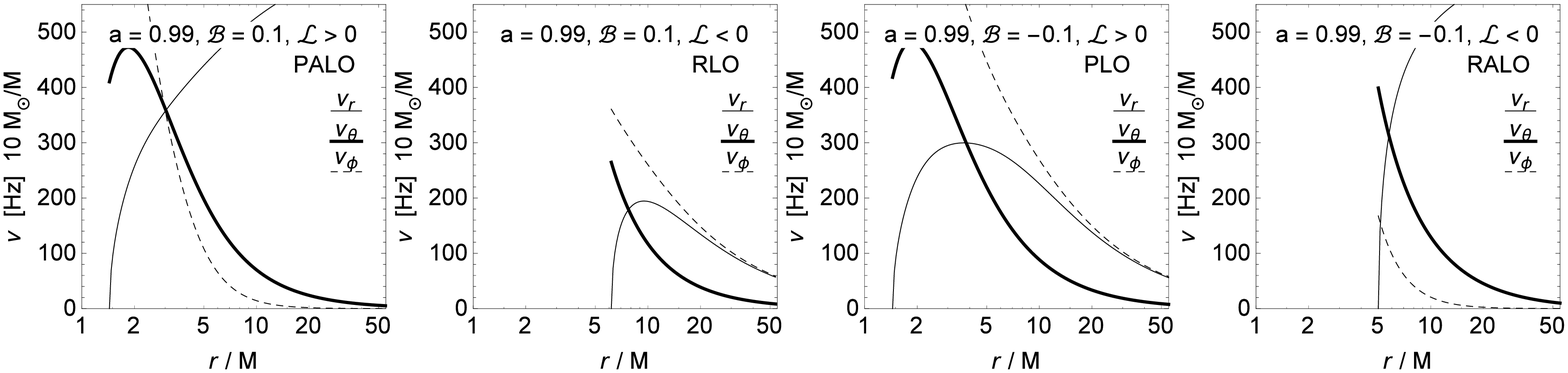}
\caption{Radial profiles of the frequencies of small harmonic oscillations $\nu_\theta, \nu_r$ and  $\nu_\phi$ of charged particle around Kerr black hole with mass $M=10 M_{\odot}$ and spin $a$ in external magnetic field measured by static distant observer.
The first row represents the influence of magnetic parameter $\BB$ in the absence of rotation (first two figures) and rotation in the nonmagnetic case (last two figures). Second row represents the Kerr black hole with $a=0.7$ with magnetic parameter $\BB=\pm0.1$. Since the differences between $\QQ=0$ and $\QQ=\QQ_{\rm W}$ cases is negligible, see discussion of eq. (\ref{diffQWandQ0a}-\ref{diffQWandQ0b}), the case $\QQ=\QQ_{\rm W}$ is considered only.
Last row represents the effect of nearly extremal  Kerr black hole with $a=0.99$ for the values of magnetic parameter $\BB=\pm0.1$.
\label{freqHZ}
} 
\end{figure*}

The expressions for the fundamental frequencies (\ref{Omtheta}), (\ref{Omradial}) and (\ref{OmKepLar}) are given in the dimensionless form. In the physical units, one needs to extend the corresponding formulas by the factor $c^3/GM$. Then the frequencies of the charged particle radial and latitudinal harmonic oscillations, and the Keplerian frequency, measured by the distant observers in Hz, are given by 
\beq
     \nu_{i} = \frac{1}{2\pi} \frac{c^3}{GM} \, \Omega_{i} [{\rm Hz}], \label{nu_rel}
\eeq
where $i\in\{r,\theta,\phi\}$. The radial profiles of the frequencies $\nu_{i}$ are shown in Fig. {\ref{freqHZ}}, the analysis of the properties of these frequencies and their relations is given in the following subsection.

\subsection{Frequencies of charged particle oscillations}


The properties of the motion of charged particles around a black hole immersed in an uniform magnetic field can be separated into four qualitatively different classes \citep{Tur-Stu-Kol:2016:PRD:}. For the particles a with positive charge one can distinguish the following orbits 

i. {\it Prograde anti-Larmor orbits} (PALO) with $\cl>0, \cb>0$. Particle is co-rotating. Magnetic field lines co-oriented with the rotation axis of the black hole. The Lorentz force is repulsive, i.e., directed outwards the black hole. 

ii. {\it Retrograde Larmor orbits} (RLO) with $\cl<0, \cb>0$. Particle is counter-rotating. Magnetic field lines co-oriented with the rotation axis of the black hole. The Lorentz force is attractive, i.e., directed towards the black hole.

iii. {\it Prograde Larmor orbits} (PLO) corresponding to $\cl>0, \cb<0$. Particle is co-rotating. Magnetic field lines counter-oriented to the rotation axis of the black hole. The Lorentz force is attractive, i.e., directed towards the black hole.

iv. {\it Retrograde anti-Larmor orbits} (RALO) corresponding to $\cl<0, \cb<0$. Particle is counter-rotating. Magnetic field lines counter-oriented with the rotation axis of the black hole. The Lorentz force is repulsive, i.e., directed outwards the black hole.

In case of the non-rotating \Schw{} black holes, PALO/RALO and RLO/PLO classes coincide. In the absence of a magnetic field, for rotating black holes, the classes PALO/PLO and RALO/RLO coincide. 

\begin{figure}
\includegraphics[width=0.554\hsize]{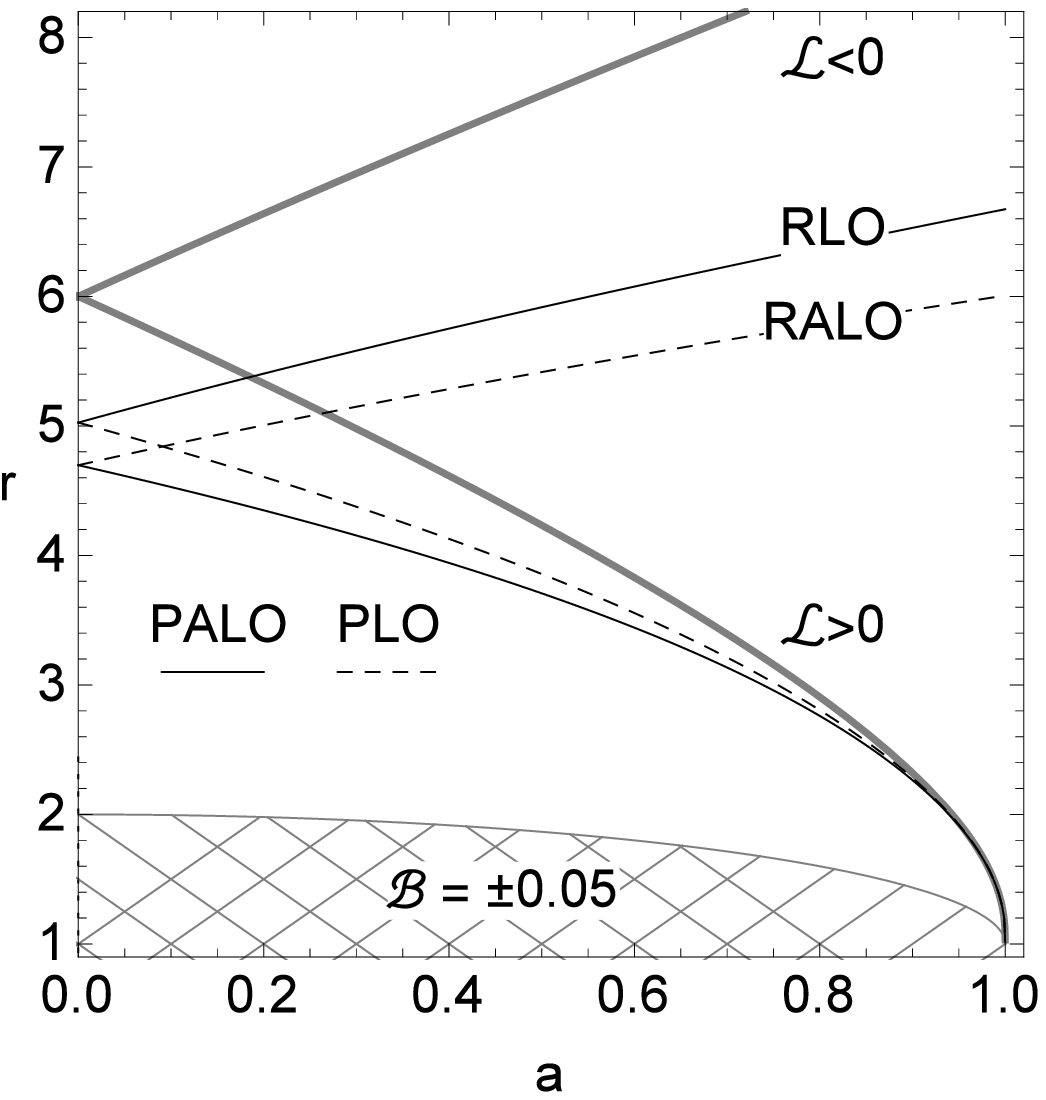}
\caption{Position of particle innermost stable circular orbit (ISCO) in rotating Kerr black hole spacetime. Co-rotating particle ISCO $\cl>0$ is shifted towards black hole horizont with increasing black hole spin $a$, while counter-rotating particle ISCO $\cl<0$ is shifted away. Non-magnetic case $\BB=0$ is plotted as thick gray curves, for positive value of magnetic parameter $\BB=0.05$ (PALO,RLO) we use solid black curve, while for negative value of magnetic parameter $\BB=-0.05$ (PLO,RALO) we use dashed black. Hatched region is area below outer horizon.
\label{figISCO}
} 
\end{figure}

Classifying orbits as given above, we plot the radial profiles of the frequencies $\nu_\theta, \nu_r$ and $\nu_\phi$ of small harmonic oscillations of charged particle measured by distant static observer in Fig. \ref{freqHZ}, for characteristic values of black hole spin $a$ and magnetic parameter $\BB$. The frequencies have been calculated using formula (\ref{nu_rel}) in case of small displacements from equatorial plane. The magnetic field presence increases the value of the radial (horizontal) frequency $\nu_r$ and the largest increase of the frequency $\nu_r$ can be observed in the PALO/RALO cases. It is also notable that $\nu_r$ does not tend to zero for large values of radial coordinate $r$; this is caused by the role of the uniform magnetic field at large distances from the black hole. The radial frequency $\nu_{r}$ vanishes at the innermost stable circular orbit (ISCO), 
\beq
\nu_{r} (r_{\rm ISCO}) = 0. 
\eeq
The position of ISCO for magnetic parameter $\BB=\pm0.05$, in dependence on black hole spin $a$, is plotted in Fig. \ref{figISCO}. The detailed analysis of the properties of the ISCO radii for different values of the magnetic field parameter has been presented in \cite{Tur-Stu-Kol:2016:PRD:}. Below the ISCO radius, the particle starts to spiral down into the black hole. The Keplerian frequency $\nu_{\phi}$ related to the orbital coordinate velocity by relation (\ref{OmKepLar}) behaves differently for different classes of the circular orbits. For PALO/RALO configurations, with repulsive Lorentz force, $\nu_{\phi}$ frequency (as well as orbital $u^\phi$ velocity \citep{Tur-Stu-Kol:2016:PRD:}, decreases with increasing the magnetic parameter $\BB$. For PLO/RLO configurations, with attractive Lorentz force, the frequency $\nu_{\phi}$  increases with increasing the magnetic parameter $\BB$. The radial profile of the vertical frequency $\nu_{\theta}$ does not have strong dependence on magnetic parameter $\BB$. Since the charged particle motion in vertical direction is not affected by the magnetic field, the Lorentz force acting on a charged particle is oriented in horizontal direction.

The frequencies in Fig. \ref{freqHZ}. have been calculated for Kerr black hole with $M=10~{M}_{\odot}$ mass. The change of the mass of the black hole leads to the change only in scale of the plots, but not their shapes due to (\ref{nu_rel}). The frequencies of oscillations depend also on the induced charge parameter $\QQ$; however, for small values of parameter $\BB$ and non-extremal spin $a$, the difference between $Q=0$ and $Q=Q_{\rm W}$ cases is negligible, see equations (\ref{diffQWandQ0a}-\ref{diffQWandQ0b}). Therefore one can use the only, e.g., $Q=Q_{\rm W}$ case, which is plotted in Fig. \ref{freqHZ}. 

In the Newtonian theory of gravity the fundamental frequencies $\nu_r, \nu_\theta$ and $\nu_\phi$ of a particle moving around central object are equal to each other 
\beq
 \nu_r = \nu_\theta = \nu_\phi = \frac{1}{2\pi} \frac{c^3}{GM} \frac{1}{r^{3/2}},
\eeq
and the elliptical trajectory is the only possible trajectory for bound orbits. For an uncharged particle orbiting \Schw{} black hole, we have $\nu_r\neq\nu_\theta=\nu_\phi$ and hence the perihelion shift of bound quasi-elliptical trajectory can be observed. 

Fundamental frequency difference in full GR theory allows construction of different HF QPOs geodesic models, where the ratio between fundamental frequencies become important. In the absence of magnetic field, the point where any two of fundamental frequencies could cross each other does not appear in the vicinity of a black hole. The frequencies coincide only in far distances from a black hole where the spacetime is asymptotically flat. But in the presence of magnetic field, the radial profiles of frequencies $\nu_r, \nu_\theta$ and $\nu_\phi$  cross each other in the vicinity of a black hole as demonstrated in Fig. \ref{freqHZ}. 
 
Concentrating our attention on the stable circular perturbations, we will further focus on small values of the magnetic parameter $\BB<1$. But it is worth to note, that for large values of $\BB>>1$, the radial profiles of fundamental frequencies in RLO and PLO motion practically ceases the change of their profiles with increasing $\BB$. Another interesting fact is that the charged particle motion in magnetized black hole background have many similarities with the "string-loop" motion which was described in \cite{Stu-Kol:2012:JCAP:,Stu-Kol:2014:PHYSR4:,Kol-Stu-Tur:2015:CLAQG:}.

\section{QPO models and influence of magnetic fields} \label{observations}

The charged particle oscillations around circular orbits, studied in the previous section, suggest interesting astrophysical application, related to LF QPOs and HF QPOs observed in many Galactic Low Mass X-Ray Binaries (LMXB) containing neutron~stars \citep{Bar-Oli-Mil:2005:MONNR:} or black holes \citep{Rem:2005:ASTRN:,Bell-etal:2012:MNRAS}. We have selected three microquasars, GRS 1915+105, XTE 1550-564 and GRO 1655-40, where the central object is a black hole candidate and for which the masses $M$ and spins $a$ are estimated, see Tab. \ref{tab1}. The limits on the values of the mass and the spin of the black hole using the spectral continuum fitting, which is independent of the QPO methods and does not include the effect of external large-scale magnetic fields, can be found in \cite{Sha-etal:2006:ApJ:,Rem-McCli:2006:ARAA:,Ste:2014:MNRAS:}. In the following subsection, we discuss the general constraint methods of black hole parameters from QPOs and the possible influence of external magnetic fields on the predictions of the rotation and mass parameters of black holes. In the later subsections we describe the general technique useful for the QPO fittings and particularly consider two most common QPO models, known as epicyclic resonance and relativistic precession models, modifying them accordingly by taking into account the effects of magnetic field.

\subsection{General properties of QPOs and constraints of black hole parameters}

The peaks of LF~QPOs with frequencies $f_{\rm low}$ and the twin peaks of the~HF~QPOs with upper $f_{\mathrm{U}}$ and lower $f_{\mathrm{L}}$ frequencies are sometimes observed in the~Fourier power spectra. In the microquasars, i.e., LMXB systems containing a black hole, the twin HF QPOs appear at the fixed frequencies that usually have nearly exact $3:2$ ratio \citep{McC-etal:2011:CLAQG:}. The observed high frequencies are close to the orbital frequency of the~marginally stable circular orbit representing the~inner edge of the accretion disks orbiting black holes; therefore, the~strong gravity effects are believed to be relevant for the  explanation of HF~QPOs \citep{Tor-etal:2005:ASTRA:}. 


\begin{figure}
\includegraphics[width=0.6\hsize]{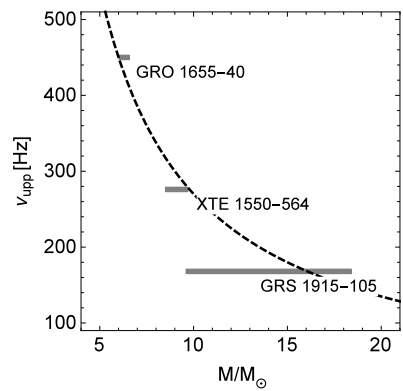}
\caption{Fitting the observed HF QPOs frequencies with the formula (\ref{simple}) plotted as dashed line. On the horizontal axis we have black hole predicted mass, on the vertical line is observed upper QPOs frequency $f_{\rm U}$.  
\label{figFF}
} 
\end{figure}

The models of twin HF QPOs involving the orbital motion of matter around black hole can be generally separated into four classes: the hot spot models (the relativistic precession model and its variations \citep{Ste-Vie:1999:PHYSRL:,Stu-Kot-Tor:2013:ASTRA:}, the tidal precession model \citep{Kos-etal:2009:ASTRA:}), resonance models \citep{Tor-etal:2005:ASTRA:,Stu-Kot-Tor:2011:ASTRA:,Stu-etal:2015:ACTA:} and disk oscillation (diskoseismic) models \citep{Rez-etal:2003:MNRAS:,Mon-Zan:2012:MNRAS:}. These models were applied to match the twin HF QPOs and the LF QPO for the microquasar GRO J1655-40 in \cite{Stu-Kol:2016:ASTRA:}. 
Of course, the models can be applied also for intermediate massive black holes \citep{Stu-Kol:2015:MNRAS:}.

\begin{figure}
\includegraphics[width=0.6\hsize]{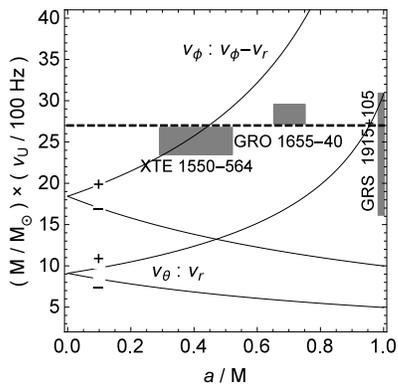}
\caption{Fitting the observed microquasar parameters with epicyclic resonance and relativistic precession models of HF QPOs (solid lines) in the absence of magnetic field. As it can be seen, none of the most common HF QPOs models is able to explain the observed HF QPOs for all the three microquasars. Plus sign is used for corotating while minus for counter-rotating orbits.
On the horizontal axis the black hole spin $a$ is given, while on the vertical axis we give black hole mass $M$ divided by observed upper HF QPOs frequency $\nu_{\rm U}$. Such mass scaling is using $\nu_{\rm U}\sim1/M$ relation (see eq. (\ref{simple})) and allow us to compare all three microquasars in one figure. 
Black boxes are predicted values of black hole mass $M$ and spin $a$, see table \ref{tab1}, the thick dashed line is given by heuristic equation (\ref{simple}).
\label{figBursa}
} 
\end{figure}

Unfortunately, none of the models recently discussed in literature, based on the frequencies of the harmonic geodesic epicyclic motion, is able to explain the HF QPOs in all three microquasars simultaneously, assuming that their central attractor is a black hole \citep{Tor-etal:2011:ASTRA:}. There is no generally accepted QPOs model for micorquasars, the observed HF QPO frequencies can be nearly fitted by the following heuristic formula (see Fig. \ref{figFF})
\beq 
\mu_{\rm U} = 2.7 \mathrm{kHz} (M_{\odot}/M). \label{simple} 
\eeq
Presented formula is useful not only for the black holes with few solar masses (including three microquasars presented in our paper), but also for objects with completely different mass magnitudes such as active galactic nuclei \citep{McCli-Rem:2004:CompactX-Sources:,Tor-etal:2005:ASTRA:,Bur:2011:POS:}. 
Existence of a simple formula (\ref{simple}) nearly fitting the observational data implies the presence of the general mechanism in black hole physics related to the generation of QPOs. Moreover, if such a mechanism exists, it has to be almost independent (or weakly dependent) on the black hole spin, since, e.g., considered three microquasars  have completely different spin estimations (see Tab. \ref{tab1}), while can be fitted using the formula (\ref{simple}), independent of the spin parameter $a$. As one can see from Fig. \ref{figBursa}, the straight dashed line corresponding to Eq.~(\ref{simple}) fits all the three microquasars with the different values of the spin parameter $a$. It is quite challenging to find a single model within which one can fit the frequencies with predicted mass $M$ and spin $a$ parameters \citep{Bur:2011:POS:}. 

As it will be demonstrated in the next subsections, the presence of external uniform magnetic field can change the situation substantially allowing us to perform the fitting of all the three microquasar sources of HF QPOs. However, one has to point out that taking into account the effects of magnetic fields can potentially affect the predictions of the black hole spin and the mass, given in Tab. \ref{tab1}. 
As it was mentioned above, the estimates of the parameters $M$ and $a$ given in the Table \ref{tab1} are based on the fitting of the continuum spectrum which does not take into account the possible interaction of charged particles with external magnetic field. While the mass of the central object can be estimated using several different methods, e.g. observing the motion of the companion star, the prediction of the black hole spin parameter is quite sensitive on the position of the inner edge of an accretion disk, which is believed to be located close to the innermost stable circular orbit (ISCO) of test particle. Particularly, in the spectral continuum models the ISCO location is one of the most important characteristics influencing the spin parameter constraints. 
However, the ISCO position for the charged particle moving around black hole significantly depends on the magnetic field, as one can see in Fig. \ref{figISCO}, or in Fig. 6 and 7 of our recent paper \citep{Tur-Stu-Kol:2016:PRD:}. For instance, the presence of such a small magnetic parameter $\BB=0.05$ in case of the Schwarzschild black holes, shifts the position of the ISCO radius considerably, decreasing it from $6 M$ (for neutral particle) to $5 M$. This occurs due to the Lorentz force acting on a charged particle, which shifts the ISCO position towards the black hole horizon. For the large values of magnetic parameter $\BB>>1$ the ISCO position can be shifted extremely close to the horizon radius of the black hole. Similar shift of the ISCO position occurs with increasing the spin parameter of the black hole. The ISCO radius coincides with the horizon radius in the extremely rotating black hole case, $a=1$. 
This implies that the uniform magnetic field can in some sense mimic the black hole spin $a$ in the sense that both increasing $a$ and $\cb$ shift ISCO to the black hole horizon for corotating particles, see Fig. \ref{figISCO}. 
Thus, the implication of a magnetic field on the spectral continuum model could lead to the new predictions of black hole spin parameter $a$, shifting it to lower values, $a_{\rm mag}<a_{\rm nomag}$, than the one we have in the 'non-magnetic' spectral continuum fitting (Table~\ref{tab1}).

The influence of strong magnetic fields (with $\BB>>1$) on the {\it profiled Fe spectral lines} has been studied in \cite{Zak-etal:2003:MNRAS} where the authors calculated the splitting of the profiled Fe spectral lines through the Zeeman effect. In case of a weak magnetic field, the modified profiled Fe spectral line originate from different angular velocity distribution inside the accretion disk, and the shift of the inner edge of the accretion disk towards the black hole horizon \citep{Fro-Sho-Tzo:2014:PRD:,Fro-Sho-Tzo:2014:JCAP:}.

\begin{table}[!ht]
\begin{center}
\begin{tabular}{c l l l}
\hline
Source & GRO 1655-40 & XTE 1550-564 & GRS 1915+105 \\
\hline \hline
$ f_{\rm U}$ [Hz] & $447${\lin}$453$ & $273${\lin}$279$ & $165${\lin}$171$ \\
$ f_{\rm L}$ [Hz] & $295${\lin}$305$ & $179${\lin}$189$ & $108${\lin}$118$ \\
$ f_{\rm low}$ [Hz] & $17.3$ & $10$ & $10$ \\
$ M/{M}_\odot $ & $6.03${\lin}$6.57$ & $8.5${\lin}$9.7$ & $9.6${\lin}$18.4$ \\
$ a $ & $0.65${\lin}$0.75$ & $0.29${\lin}$0.52$ & $0.98${\lin}$1$ \\
\hline
\end{tabular}
\caption{\label{tab1}
Observed twin HF QPO data for three microquasars and the restrictions on mass and spin of the black holes located in them, based on measurements independent of the HF QPO measurements given by the optical measerument for mass estimates and by the spectral continuum fitting for spin estimates \citep{Sha-etal:2006:ApJ:,Rem-McCli:2006:ARAA:}. 
Note that in the GRS~1915+105 microquasar more HF QPOs are observed \cite{Bel-Alt:2013:MNRAS}. 
Here we concentrate attention to the pair of HF QPOs demonstrating the frequency ratio 3:2, common with those observed in other two microquasars.
} 
\end{center}
\end{table}

 \subsection{Resonant radii and the fitting technique}

The HF QPOs come in pair of two peaks with upper $f_{\mathrm{U}}$ and lower $f_{\mathrm{L}}$ frequencies in the timing spectra. For the QPOs from black hole microquasars given in Tab. \ref{tab1}, the frequency ratios $f_{\mathrm{U}}:f_{\mathrm{L}}$ are very close to the fraction $3:2$. Observation of this effect in different non-linear systems indicates the existence of the resonances between two modes of oscillations. In case of geodesic QPO models, the observed frequencies are associated with the different linear combinations of the particle fundamental frequencies $\nu_{r}, \nu_{\theta}$ and $\nu_{\phi}$. In the presence of magnetic field the upper and lower frequencies of HF QPOs are the functions of magnetic parameter $\BB$, black hole mass $M$, spin $a$ and the resonance position $r$,
\beq
 \nu_{\mathrm{U}} = \nu_{\mathrm{U}}(r,M,a,\BB), \quad \nu_{\mathrm{L}} = \nu_{\mathrm{L}}(r,M,a,\BB). \label{ffUL}
\eeq
It is worth to note that the frequencies $\nu_{\mathrm{U}}$ and $\nu_{\mathrm{L}}$ are inversely proportional to the mass $M$ of a black hole, while the dependence of frequencies on the spin $a$ and magnetic field $\BB$ is more complicated and hidden inside $\Omega_{r},\Omega_{\theta},\Omega_{\phi}$ functions, as given by equation (\ref{nu_rel}).

The resonant models of the twin HF QPOs assume a particular parametric or non-linear forced resonance of the oscillatory modes of the accretion disk \citep{Ali-Gal:1981:GRG:}. In order to fit the frequencies observed in HF QPOs with the black hole parameters, one needs first to calculate the so called resonant radii $r_{3:2}$ ($r_{2:3}$)
\beq
\nu_{\mathrm{U}}(r_{3:2}):\nu_{\mathrm{L}}(r_{3:2})=3:2, \quad 
\nu_{\mathrm{U}}(r_{2:3}):\nu_{\mathrm{L}}(r_{2:3})=2:3. \label{rezrad}
\eeq
Resonant radii $r_{3:2}$ (or $r_{2:3}$) in general case are given as the numerical solution of higher order polynomial in $r$, for given values of spin $a$ and magnetic field $\BB$ parameters. Since the equation (\ref{rezrad}) is independent of the black hole mass explicitly, the resonant radius solution also has no explicit dependence on the black hole mass and techniques introduced in \cite{Stu-Kot-Tor:2013:ASTRA:} can be used. Substituting the resonance radius into the equation (\ref{ffUL}), we get the frequencies $\nu_{\mathrm{U}}$ and $\nu_{\mathrm{L}}$ in terms of the black hole mass, spin and magnetic field.

The inverse dependence of the frequencies on the black hole mass allows us to write the mass $M$ in terms of frequency, spin and magnetic field. The example of this dependence in the absence of magnetic field is shown in the Fig. \ref{figBursa}. As it can be seen in the Fig. \ref{figBursa}, in the absence of magnetic field, the calculated frequencies are quite sensitive on the black hole spin $a$, while the simple heuristic formula (\ref{simple}) given by dashed line is not dependent on black hole spin $a$.


In the following, we will examine magnetic field contribution to the two most relevant geodesic QPO models: epicyclic resonance  and relativistic precession models.

\begin{figure*}
\includegraphics[width=\hsize]{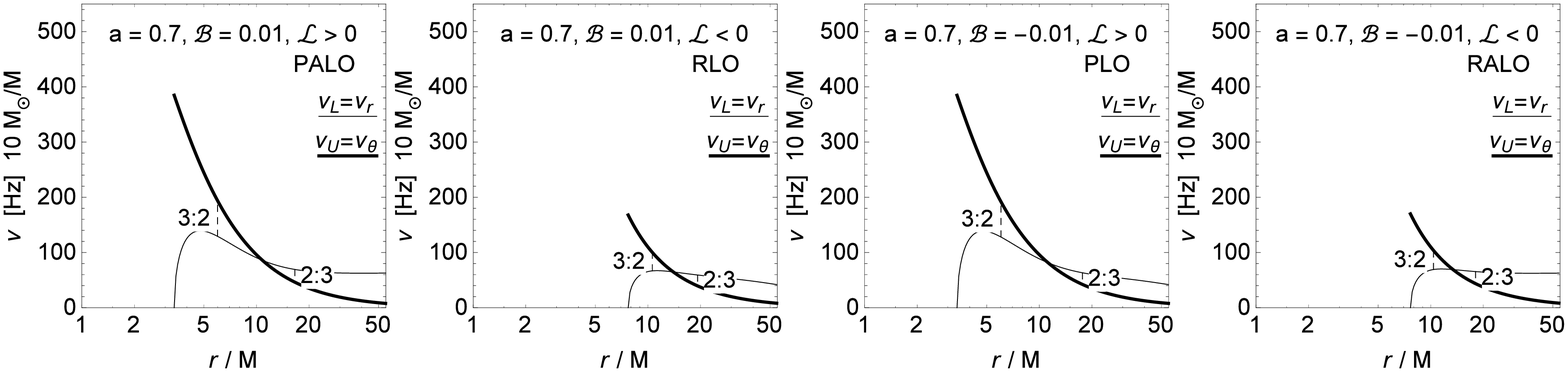}
\includegraphics[width=\hsize]{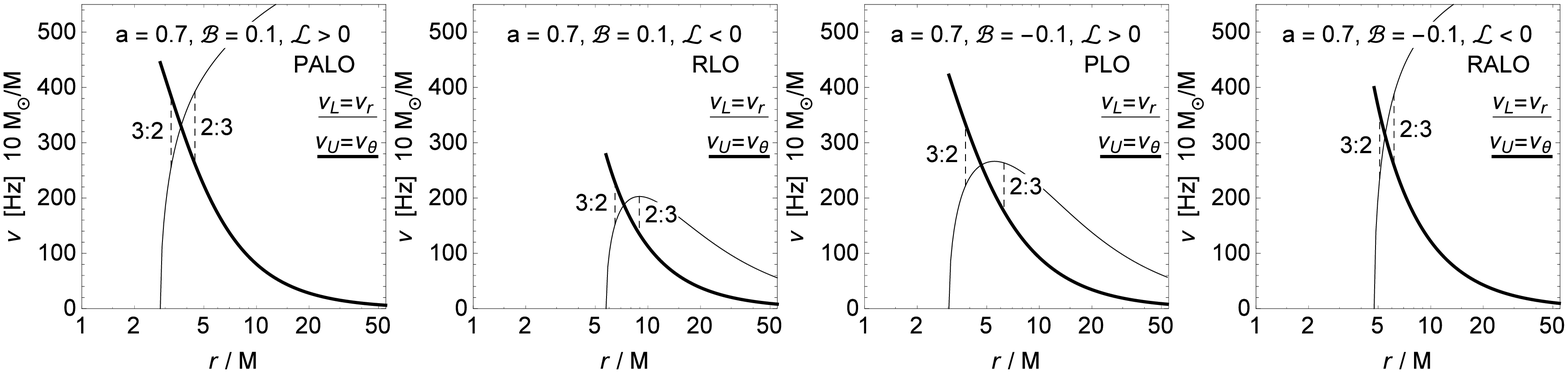}
\caption{
Radial profiles of upper $\nu_{\rm U}=\nu_{\theta}$ and lower $\nu_{\rm L}=\nu_{\rm r}$ frequencies in HF QPOs {\bf ER model}. 
The location of both resonance radii $3:2$ and $2:3$ are going closer to ISCO radius with increasing the magnetic parameter $\BB$. $2:3$ resonance is possible only in the presence of magnetic field. For $\BB=0$ and $a=0$ the resonance radii are placed at $r_{3:2}\doteq{10.7}$ while $r_{2:3}\rightarrow\infty$. 
Comparing the first row (plotted for $\BB=0.01$) with the second row (plotted for $\BB=0.1$) we can see, that upper $\nu_{\rm U}$ and lower $\nu_{\rm L}$ resonance frequencies (at resonance radii) are increasing with increasing the magnetic parameter $\BB$. The lower frequency $\nu_{\rm L}$ is more influenced by magnetic field presence than the upper one $\nu_{\rm L}$, but this appears due to the influence of the Lorentz force in radial direction.
\label{figER}
}
\end{figure*}
\begin{figure*}
\includegraphics[width=\hsize]{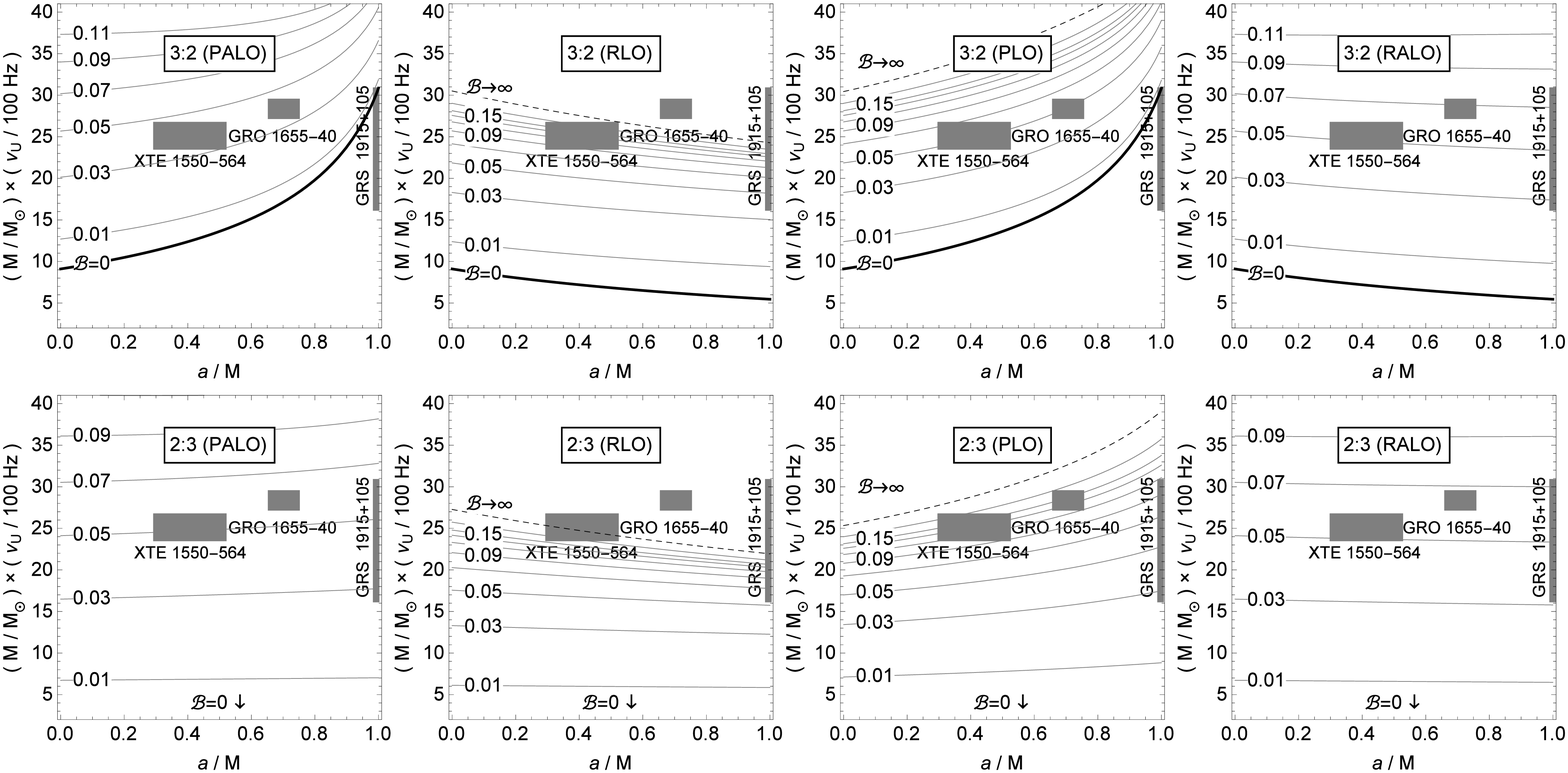}
\caption{
Fitting of three microquasar sources with HF QPOs {\bf ER model} of charged particle oscillations with the effect of magnetic field. On the horizontal axis the black hole spin $a$ is given, while on the vertical axis we give black hole mass $M$ divided by observed upper HF QPOs frequency $\mu_{\rm U}$. Gray boxes are the values of black hole mass $M$ and spin $a$ estimated from the observations, see table \ref{tab1}. Thin lines are given by equation (\ref{simple}) and (\ref{ERmodel}) for different values of magnetic parameter $\BB$. The non-magnetic case $\BB=0$ is represented by thick lines.
There could exist both 3:2 and 2:3 resonances and four different combination of magnetic field / black hole spin configurations, namely PALO, RLO, PLO and RALO, giving thus eight different cases in total. Co-rotating particle model (PALO and PLO) can fit the microquasar sources quite well, especially in the case of 2:3 resonances, where one can fit all three microquasar. Counter-rotating particle model (RLO,RALO) are successful in fitting for RALO case only, RLO case is completely unable to fit GRO 1655-40 source for any values of $\BB$, while source GRS 1915+105 can be fitted by RLO case only partially.
\label{ERmodel2}
}
\end{figure*}

\subsection{Epicyclic resonance (ER) model}

The epicyclic resonance (ER) models \citep{Abr-Klu:2001:AA:,Tor-etal:2005:ASTRA:} consider resonance of axisymmetric oscillation modes of accretion discs.  The frequency commensurability is thus crucial ingredient of the resonant models, and a particular case of this commensurability occurs for the parametric (internal) resonant phenomena that become strongest in the case of the $3:2$ frequency ratio \citep{Tor-etal:2005:ASTRA:,Stu-Kot-Tor:2013:ASTRA:}. The simplest variant of the resonant model is the resonant epicyclic model where the two resonant modes are identified with the radial and vertical epicyclic frequencies \citep{Tor-etal:2005:ASTRA:}. 
\beq
 \nu_{\rm U}=\nu_{\theta}, \quad \nu_{\rm L}=\nu_{\rm r} \label{ERmodel}
\eeq
Radial profiles for upper $\nu_{\rm U}$ and lower $\nu_{\rm L}$ frequencies for ER model are given in Fig. \ref{figER}. In the non-magnetic case $\BB=0$ (geodesic motion in Kerr spacetime), we have $\nu_{\theta}(r)>\nu_{r}(r)$ for any value of spin parameter $a$. For the magnetic case $\BB\neq0$, the behaviour of the radial profiles of the $\nu_{\rm U}$ and $\nu_{\rm L}$ frequencies become more complicated, there exist two separate regions: for small values of radial coordinate $r_{\rm ISCO}<r<r_{1:1}$ we have $\nu_{\rm U}(r)>\nu_{\rm L}(r)$; for larger values of radial coordinate $r_{1:1}<r<\infty$, we have $\nu_{\rm U}(r)<\nu_{\rm L}(r)$. In the absence of magnetic field  $\BB=0$ the radii $r_{1:1}$ where two frequencies coincide ($\nu_{\rm U}(r)=\nu_{\rm L}(r)$) is located at  infinity $r\rightarrow\infty$.
For magnetic case $\BB\neq0$, there exist two resonant radii $r_{3:2}$ and $r_{2:3}$ where one can produce the observed frequency ratios according to Tab. \ref{tab1}. Thus the radii having importance for our study are lined up in the following way: $r_{\rm ISCO}<r_{3:2}<r_{1:1}<r_{2:3}$.

Fits of the three microquasar sources, HF QPOs modified ER model of charged particle oscillations in the field of magnetized black holes, are given in Fig. \ref{ERmodel2}. Four different configurations of orbits (PALO, RLO, PLO and RALO) for both 3:2 and 2:3 resonances are given. In absence of the magnetic field, the ER model is not able to explain the observed HF QPOs in all the three microquasars, however, in the magnetic case it is possible - see magnetic field parameter $\BB$ estimations in Tab. \ref{tab2}. The co-rotating particle model (PALO,PLO) can fit the microquasar sources quite well, especially for 2:3 resonances, where we can fit all three microquasars with one value of magnetic field parameter $\BB$. Counter-rotating particle model (RLO,RALO) is successful in fitting for the RALO case only, the RLO case is completely falsified to explain the GRO 1655-40 source for any value of $\BB$, while the GRS 1915+105 source can be fitted by the RLO case only partially.

Increasing the magnetic field parameter $\BB$, the charged particle frequency will change. For the RLO and PLO cases, such changes become smaller and smaller with magnetic field $\BB$ increasing - charged particle dynamics in magnetic field approaches the "string-loop" limit \citep{Kol-Stu-Tur:2015:CLAQG:}. Moreover, when $\BB$ is increased, the predicted frequencies become more independent on the black hole spin $a$ - frequencies of charged particle oscillation become more influenced by the magnetic field than gravity for large $\BB$. The existence of new 2:3 resonance radii, at large radii, is also caused by the magnetic field presence. More detailed consequences of the fitting and discussion of results is given is Section \ref{kecy}. 

\begin{table}[!ht]
\begin{center}
\begin{tabular}{l l l l}
\hline
				 & GRO 1655-40 		& XTE 1550-564 	 & GRS 1915+105 \\
\hline \hline
3:2 PALO & 0.03{\lin}0.05 & 0.03{\lin}0.05 & $\nexists${\lin}0.01 \\
3:2 RLO  & $\nexists$ 		& 0.08{\lin}1+ 	 & 0.04{\lin}$\nexists$ \\
3:2 PLO  & 0.00{\lin}0.04 & 0.04{\lin}0.08 & $\nexists${\lin}0.01 \\
3:2 RALO & 0.06{\lin}0.07 & 0.05{\lin}0.06 & 0.03{\lin}0.08  \\
\hline
2:3 PALO & 0.05{\lin}0.06	& 0.05{\lin}0.06 & 0.03{\lin}0.06	  \\
2:3 RLO  & $\nexists$ 	& 0.15{\lin}$\nexists$ & 0.05{\lin}$\nexists$	  \\
2:3 PLO  & 0.11{\lin}0.16 & 0.08{\lin}0.16 & 0.03{\lin}0.12   \\
2:3 RALO & 0.06{\lin}0.07 & 0.05{\lin}0.06 & 0.03{\lin}0.07  \\
\hline
\end{tabular}
\caption{Constraints of the magnetic parameter $\BB$ for three microquasars obtained by the fitting of HF QPOs with ER model.
\label{tab2}
} 
\end{center}
\end{table}


\begin{figure*}
\includegraphics[width=\hsize]{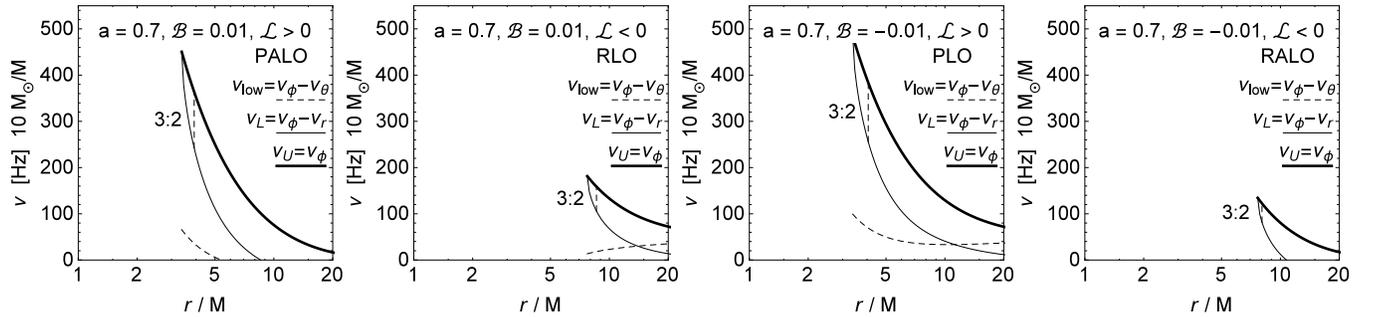}
\caption{
Radial profiles of upper $\nu_{\rm U}=\nu_{\phi}$ and lower $\nu_{\rm L}=\nu_{\phi}-\nu_{\rm r}$ frequencies in HF QPOs {\bf RP model}. Resonance radius $3:2$ is given; location is going closer to ISCO with the increasing of the magnetic parameter $\BB$. 
Comparing the first row (plotted for $\BB=0.01$) with the second row (plotted for $\BB=0.1$) we can see, that upper $\nu_{\rm U}$ and lower $\nu_{\rm L}$ rezonance frequencies (at resonance radius) are increasing with magnetic parameter $\BB$ increase.
\label{figRP1}
}
\end{figure*}
\begin{figure*}
\includegraphics[width=\hsize]{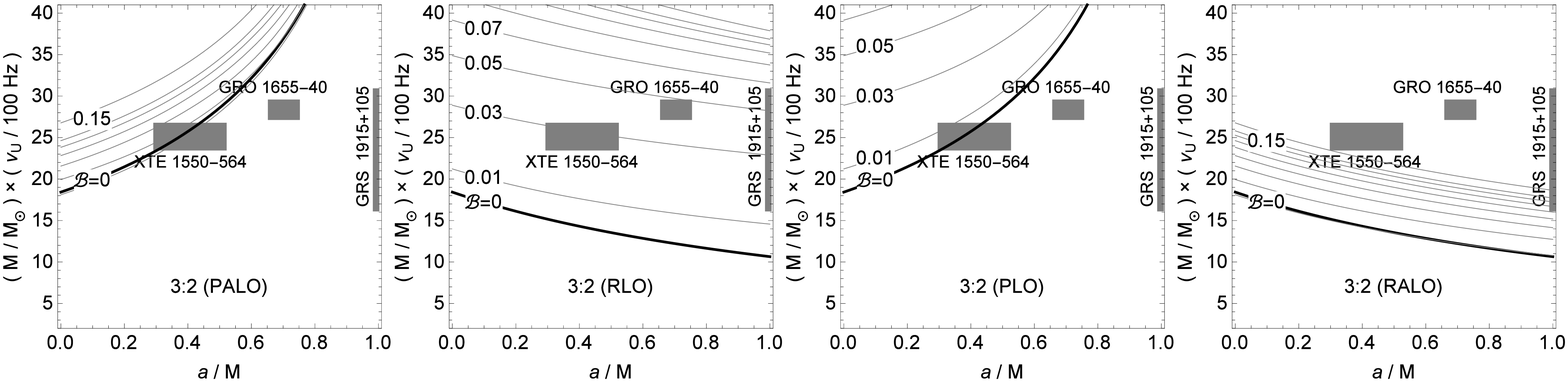}
\caption{
Fits of three microquasar sources, using HF QPOs {\bf RP model} of charged particle oscillations with magnetic field. On the horizontal axis the black hole spin $a$ is given, while on the vertical axis we give black hole mass $M$ divided by observed upper HF QPOs frequency $\mu_{\rm U}$. Gray boxes are predicted values of black hole mass $M$ and spin $a$, see table \ref{tab1}. The lines are given by equation (\ref{simple}) and (\ref{RPmodel}) for different values of magnetic field parameter $\BB$, for non-magnetic case $\BB=0$ the lines are thicker.
Only counter-rotating particle model RLO is successful in fitting all three sources.
\label{figRP2}
}
\end{figure*}

\subsection{Relativistic precession (RP) model for HF QPOs}

Relativistic precession (RP) model was introduced in order to explain the behaviour of the QPOs observed in X-ray neutron star and black hole binaries \citep{Ste-Vie:1998:ApJ:,Ste-Vie-Mor:1999:ApJ:,Mot-etal:2014a:MNRAS:}. The RP model belongs to the class of geodesic QPO models, with the frequencies of $\nu_{\rm U},\nu_{\rm L}$ for HF QPOs defined as
\beq
 \nu_{\rm U}=\nu_{\phi}, \quad \nu_{\rm L}=\nu_{\phi}-\nu_{\rm r}. \label{RPmodel}
\eeq
Radial profiles of the upper $\nu_{\rm U}$ and lower $\nu_{\rm L}$ frequencies for the RP model, and position of the related $3:2$ resonance radii, are given in Fig. \ref{figRP1}. 

In the absence of magnetic field ($\BB=0$), the RP model gives $\nu_{\rm \phi}>\nu_{\rm r}$. When magnetic field is taken into account, the limit of the RP model for PALO/RALO configurations corresponds to the emergence of $\nu_{\rm r}=\nu_{\rm \phi}$ radii. So for large $\BB$ there exist radii at which $\nu_{\rm \phi}<\nu_{\rm r}$ and the lower frequency $\nu_{\rm L}$ given by (\ref{RPmodel}) in RP model will become negative and has to be redefined.
For RLO/PLO cases, we have $\nu_{\rm \phi}>\nu_{\rm r}$  for any value of $\BB$ parameter and the RP modes work properly. 

\begin{figure*}
\includegraphics[width=0.85\hsize]{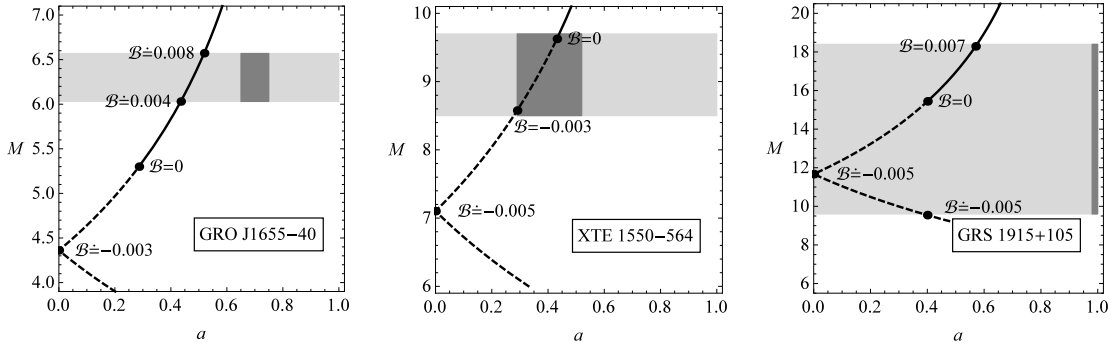}
\caption{
HF and LF QPOs RP model predictions of black hole mass $M$ and spin $a$ for various values of magnetic filed $\BB$ are given by thick black curve. For positive values of $\BB$ parameter the curve is solid, for nagative values $\BB<0$ the curve is dashed. Any point o the curve is solution of three equations (\ref{RPmodel3}) for three parameters $M, a, \BB$. The dark grey area denotes restrictions on $M$ and $a$ form from spectral fits, light gray denotes only mass $M$ restictions.
\label{fig09}
} 
\end{figure*}

The RP HF QPOs model modified by the presence of the magnetic field is able to match all observed HF frequencies with estimated black hole mass $M$ and spin $a$, see Tab. \ref{tab1}., only in the 3:2 RLO case, see Fig. \ref{fig09}. In the PALO/RALO cases, the RP model will not work properly for large values of magnetic parameter with $\BB>0.1$, while for PLO case the magnetic field in the RP model increases the frequencies to larger than observed values. In the RLO scenario, the counter-rotating disk around central black hole is influenced by the magnetic field with strength $\BB\sim0.003$. Values of $\BB$ for every individual microquasars are given in Tab. \ref{tabRP}.
  
\begin{table}[!ht]
\begin{center}
\begin{tabular}{l l l l}
\hline
				 & GRO 1655-40 			& XTE 1550-564 		 & GRS 1915+105 \\
\hline \hline
3:2 RLO  & 0.022{\lin}0.035 & 0.039{\lin}0.051 & 0.013{\lin}0.065 \\
\hline
\end{tabular}
\caption{Constraints of the magnetic parameter $\BB$ for the three microquasars, obtained by the RP model of twin HF QPOs. \label{tabRP} } 
\end{center}
\end{table}

\subsection{RP model assuming simultaneous observations of HF and LF QPOs}

The simultaneous observation of the twin HF QPOs and the LF QPOs (see Tab.\ref{tab1}) in all three considered microquasars enables us to obtain the stringent restrictions on the mass and dimensionless spin of the central black hole, if we assume that all of these QPOs arise at a given radius of the accretion disc \citep{Mot-etal:2014a:MNRAS:}. RP model for simultaneous HF and LF QPOs has frequencies $\nu_{\rm U},\nu_{\rm L}$ defined as
\beq
 \nu_{\rm U}=\nu_{\phi}, \quad \nu_{\rm L}=\nu_{\phi}-\nu_{\rm r}, \qquad \nu_{\rm low}=\nu_{\phi}-\nu_{\theta}. \label{RPmodel2}
\eeq
Now, one can use the simultaneously observed values of HF and LF QPOs $f_{\rm U}, f_{\rm L}, f_{\rm low}$ frequencies and identify them with formula for RP model (\ref{RPmodel2}) through three equations
\bea
 && f_{\rm U}=\nu_{\rm U}(M,a,\BB), \quad f_{\rm L}=\nu_{\rm L}(M,a,\BB), \nonumber \\  
 && f_{\rm low}=\nu_{\rm low}(M,a,\BB) \label{RPmodel3}
\eea
for three independent parameters - black hole mass $M$, spin $a$ and magnetic parameter $\BB$. For different values of the magnetic parameter $\BB$, different combinations of the mass $M$ and spin $a$ can be obtained, see results in Fig. \ref{fig09}. Only for corotating particles (PALO $\BB>0$ and PLO $\BB<0$) the real solution of Eq. (\ref{RPmodel3}) can be found. 

However, the restrictions on the black hole mass $M$ and spin $a$ given by the particle oscillations in the absence of magnetic field, $\BB=0$, are in contradiction with the values obtained from the spectral fits for GRO J1655-40 and GRS 1915+105 source \citep{Stu-Kol:2016:APJ:}. For $\BB=0$, only the XTE 1550-564 source fits mass $M$ and spin $a$ restrictions from Tab. \ref{tab1}. Note that in accord with the idea of \cite{Stu-Kol:2016:APJ:}, we could remove the discrepancy with fitting to the limits on the BH spin $a$ given by the spectral measurements by breaking the assumption that the twin HF QPOs and the LF QPO arise at the same radius.

Taking magnetic field $\BB$ into account, we are able to fit mass $M$ of individual central object given in Tab. \ref{tab1}. Then the spin $a$ and magnetic field $\BB$ can be estimated using (\ref{RPmodel3}). The results are summarised in Tab. \ref{tabRPLF}, where we can observe that the RP model for HF and LF QPOs with magnetic field is predicting lower values of the spins $a$ than the predictions due to the spectral fits.

\begin{table}[!ht]
\begin{center}
\begin{tabular}{c c c c}
\hline
 & GRO 1655-40 & XTE 1550-564 & GRS 1915+105 \\
\hline \hline
$\BB$ & 0 	 & 0						& 0 \\
$M$ &	 5.3	 & 		9.6 			&  15.4 \\
$a$ &	 0.29	 &  	0.43 			&  0.40 \\
\hline
$\BB$	 & 0.004{\lin}0.008 & -0.003{\lin}0.000 & -0.005{\lin}0.007 \\
$ M $ & $6.03${\lin}$6.57$ & $8.5${\lin}$9.7$ & $9.6${\lin}$18.4$ \\
$a$		 &	0.43{\lin}0.52  &   0.28{\lin}0.44 & 0{\lin}0.56 \\
\hline
\end{tabular}
\caption{Black hole mass $M$ and spin $a$ restrictions for three microquasars given by RP model with LF QPOs. We used HF and LF QPOs frequencies from Tab.~\ref{tab1}. First three rows are predictions for nonmagnetic $\BB=0$ case, while the second there rows are spin $a$ and magnetic field $\BB$ predictions for observed mass $M$ restrictions from Tab.~\ref{tab1}.
\label{tabRPLF}
}
\end{center}
\end{table}


\section{Discussion} \label{kecy}

We have studied the influence of an external weak magnetic field, approximated as asymptotically uniform with field lines oriented along the black hole rotation axis, on the motion of charged particle in its relation to the explanation of QPOs observed in microquasars. We have shown that the effect of even weak magnetic field on the motion of charged particles cannot be neglected and moreover, in the oscillatory phenomena of charged particles the influence of magnetic field is dominant over the influence of the black hole rotation. We have demonstrated that in order to fit the observational frequencies of HF QPOs with the epicyclic resonance model, the effect of external magnetic field can be sufficient. 

\begin{table}[!ht]
\begin{center}
\begin{tabular}{c@{\qquad} c@{\quad} c@{\quad} c@{\quad} c } 
\hline
  & electron & proton & Fe+ & charged dust \\
\hline
\hline
$\BB = 0.004$ & $10^{-5}~{\rm{G}}$ & $ 0.02~{\rm{G}} $ & $ 1~{\rm{G}}$ &  $ 10^8~{\rm{G}}$\\
\hline
\end{tabular}
\caption{Magnitudes of magnetic fields $B$ in Gauss units for magnetic parameter $\BB=0.004$ and stellar mass black hole $M={10~{M}_{\odot}}$. Test particles with different specific charges $q/m$ ratios are assumed - from electron, proton and partially ionized Ferrum atom (one electron lost) to charged dust grain (one electron lost, $m=2\times10^{-18}$~kg).
\label{tabX}
} 
\end{center}
\end{table}

In this article we assume the magnetic field to be uniform. Real magnetic fields around microquasar black holes and their accretion disks are far away from being completely regular and uniform, the Wald uniform magnetic field solution is used as a useful approximation. In the case of QPOs models, it is enough to assume uniformity of the magnetic field in the region where the oscillatory motion occurs. Fitting of the observed QPOs with the charged particle oscillations put limits on the values of the magnetic parameter $\BB$, which however contains, together with the field strength, also the specific charge of the oscillating test particles. This implies that in order to make proper estimation of the magnetic field values in the vicinity of particular microquasars, one needs to identify first the type of oscillating matter. In our approach, the "charged particle" can represent matter ranging from electron to some charged inhomogeneity orbiting in the innermost region of the accretion disk. The specific charges $q/m$ for any of such structures will then range from the electron maximum to zero. Recalling the physical constants in the dimensionless magnetic parameter as $\BB = |q| B G M/(2 m c^4)$, we get the magnetic field strength in Gauss
\beq 
B = \frac{2 m c^4 \BB}{q G M} ~ [{\rm G}], \label{MFgauss} 
\eeq
where the quantities are given in CGS units, see Tab. \ref{tabX}.

Assuming the oscillating particles to be electrons (as the main candidate source of synchrotron radiation producing X-rays), we get the estimations of the strength of magnetic field in the regions around the three microquasars as given in Table \ref{tab3}. Surprisingly, the range of the magnetic field magnitude estimates coincides with the measurements of the strength of large-scale Galactic magnetic fields.

\begin{table}[!ht]
\begin{center}
\begin{tabular}{l l l l}
\hline
				 & GRO 1655-40 	& XTE 1550-564 	 & GRS 1915+105 \\
\hline \hline
\multicolumn{3}{l}{ER model}\\
PALO & 11.5{\lin}23.0   & 8.2{\lin}16.3 		& 0{\lin}14.5 \\
RLO  & $\nexists$  		  & 21.8{\lin}$\infty$& 9.6{\lin}$\infty$ \\
PLO  & 0{\lin}61.4			& 10.9{\lin}43.6  	& 0{\lin} 28.9  \\
RALO & 23.1{\lin}26.9   & 13.6{\lin}16.3 		& 7.2{\lin}19.3  \\
\hline
\multicolumn{3}{l}{RP model for HF QPOs only}\\
RLO & 8.4{\lin}13.4 & 10.6{\lin}13.9 & 3.1{\lin}15.7 \\
\hline
\multicolumn{3}{l}{RP model for HF and LF QPOs}\\
& 1.5{\lin}3.1 & 0.8{\lin} 0.8 & 0.0{\lin}1.7 \\
\hline
\end{tabular}
\caption{Magnitude of magnetic filed for three microquasars, obtained by the fitting of the QPOs frequencies with the frequencies of the oscillations of an electron. We use combined values for ER model 3:2 and 2:3 resonances (Tab. \ref{tab2}) and also values for RP model (Tab. \ref{tabRP}. and \ref{tabRPLF}.). All numbers in the table are in units of $10^{-5}$ Gauss.
\label{tab3}
} 
\end{center}
\end{table}

The values of magnetic parameter $\BB$ sufficient to fit the observational frequencies of different sources are given in Tab. \ref{tab2}. In all cases, one can predict the magnitude of magnetic field parameter to be of the order of $\BB\sim 10^{-2}$. Assuming that the QPOs appear due to the oscillations of electrons inside accretion disk one can estimate the magnetic field strength for a particular microquasar with the help of Eq.(\ref{MFgauss}) and Tables \ref{tab1} and \ref{tab2}. The estimates of the strength of a magnetic field in the vicinity of the three microquasars are given in Table \ref{tab3}. The different class of orbits can correspond to the different configurations of an accretion disk (corotating or counterrotating) and different alignments of the magnetic field lines. As one can see from Table \ref{tab3}, the estimated values of the magnetic field in all cases for all microquasars with different configurations of accretion disks are of the same order, $10^{-5} \div 10^{-4}$ Gauss which suggests the interpretation of the origin of these fields as the Galactic magnetic field. There are many indications that the magnetic field must be present in the interstellar medium, and our estimates are in accord with these observations. There were also several attempts to estimate the strength of magnetic field in the vicinity of the three microquasars considered in the present paper. The estimates were using different methods for each source, independently on the QPO models. Here are few of them.

Magnitude of the equipartition magnetic field around the microquasar GRS 1915+105, has been estimated to be $\sim 8$~G in \cite{Fender-etal:1997:MNRAS:}, being based on the assumption that the repeated radio events are synchrotron in origin.

Fitting the flux data of the X-ray jets from the microquasar XTE J1550-564, with the trans-relativistic external shock model based on the analogy with the gamma-ray burst remnants \citep{Wan-Dai-Lu:2005:APSS:}, the authors suggested the magnetic field strength of the value $5\times10^{-4}$~G. Similar values have been confirmed in \cite{Hao-Zha:2009:APJ:} from the observation of both jets of XTE J1550-564.

For the microquasar GRO J1655-40, the magnetic field strength has been estimated in \cite{Gre-Pre-Poh:1995:AAP:}, using the synchrotron-self-Compton model where the strength is obtained to be in the range between $0.05$~G and $49$~G, depending on the parameters of the model.

If one assumes the HF QPOs to be produced due to the oscillations of heavier than electrons particles, the estimates of magnetic field strength in our model will increase accordingly. For protons surrounding a black hole with $10~M_{\odot}$, the magnetic field strength with average value of the magnetic parameter $\BB=0.05$ gives a value of 0.2~G. Some QPOs were detected in the modulations of an iron line of spectra. Thus, if one assumes that the QPOs come from the oscillations of iron atoms, one can get the magnetic field of 12~G, for the black hole with $10~M_{\odot}$ and equipartition magnetic parameter $\BB=0.05$. The physical interpretation of the X-ray emission from heavy particles, such as ions, can be realized by considering,  e.g., the nuclear-Compton scattering processes. As one can see, the estimates of the strength of the magnetic field in our models do not disprove the independent magnetic field measurements. 
In the case of LF QPO, which was simultaneously observed with HF QPOs, the fitting by the RP model gives the average magnetic parameter $\BB=0.005$ for all sources. Using similar estimates way, as described above we get the strength of a magnetic field of an order of $(1-10)\times10^{-6}$~G assuming the synchrotron emission to be produced by electrons. 

There are still some open questions and unresolved issues in the presented magnetic HF QPOs model. QPOs frequencies $f_{\rm U}, f_{\rm L}$ observed in microquasars are notorious to be constant through time. They remain constant for different accretion disk regimes, where the accretion rates $\dot{M}$ are assumed to be different. For magnetic HF QPOs models this means that the external uniform magnetic field (or et last $\BB$ parameter) must also be constant through time. The stability of HF QPOs frequencies observed for black hole binaries, become apparent with comparison of HF QPOs observed in neutron star binaries - the neutron star HF QPO frequencies are substantially changing in time. Maybe, this tells something about the structure of magnetic field around neutron star/black hole objects. For neutron stars the magnetic field is assumed to be much stronger and complicated, having dipole character. On the other hand, our results indicate that for the black hole binaries the Galactic magnetic field could be relevant that can be expected nearly constant on year timescales relevant for QPO observations.

\section{Conclusions}

The oscillatory frequencies of charged particles in black hole surroundings filled with asymptotically uniform magnetic field can be well related to the frequencies of QPOs observed in the microquasars GRS~1915+105, XTE~1550-564, GRO~1655-40. We can summarise main results for magnetic ER and RP models as follows:
\begin{itemize}
\item Influence of magnetic field $\BB$ on charged particle oscillatory motion can mimic the influence of the black hole rotation $a$ for co-rotating particles - the ISCO can be shifted towards to the black hole horizon by magnetic field presence, and charged particle oscillatory frequencies can be increased due to the magnetic field, as shown in Fig. \ref{freqHZ}.
\item Frequencies given by the magnetized versions of the QPOs geodesic models are dependent only slightly on the black hole spin $a$ - the effect of weak magnetic field on the charged particle oscillations is stronger than the corresponding contribution of the black hole rotation. In the magnetized QPOs model, we obtained the formula $\nu\sim{1/M}$, similar to the Eq. (\ref{simple}), and hence we can fit data from all the three considered microquasars with a single QPOs model.
\item Contrary to the neutral particle motion, the charged particle motion around black hole immersed into magnetic field is driven by non-linear equations of motion. Without non-linearity there is no resonance between $\nu_{U}$ and $\nu_{L}$ modes of oscillation. Magnetized HF QPOs models will give not only the possibility to fit the values black hole spin $a$ and $M$, but provide potentially also explanation of the asumed resonances. 
\item It was demonstrated that both the magnetized versions of the standard ER and RP models of twin HF QPOs can explain the observationally fixed data from the three microquasars, GRS 1915+105, XTE 1550-564 and GRO 1655-40, if we assume nearly equal (of the same order) magnitude of the external (near-uniform) magnetic field, and limits on the black hole mass $M$ (and dimensional spin $a$) given by independent optical (spectral continuum) measurements. The assumed magnetic field magnitude is in agreement with estimates of the Galactic magnetic field for the electron case.
\item The magnetic ER model admits both the co-rotating and counter-rotating disks, while the RP model admits counter-rotating disks only. Moreover, in the RP model containing naturally also the explanation of the simultaneously observed LF QPO, we have found a contradiction with the limits on spin in the microquasars GRS 1915+105 and GRO 1655-40, if the creation of both HF QPOs and the related LF QPOs occurs at the same radius. 
As proposed in \cite{Stu-Kol:2016:APJ:}, this controversy could be solved by breaking assumption of common location of the twin HF QPOs and LF QPOs. Problem of lower predicted black hole spins $a$, given by the RP model with the LF QPOs, could be also solved, assuming magnetic corrections to the spectral continuum model, which can decrease limits the on the black hole spin.
\item In the magnetized RE or RP models related to the twin HF QPOs, the uniform magnetic field is necessary ingredient only in the vicinity of the black hole horizon where the epicyclic oscillations have to occur. Therefore, the state of the accretion disk should be relatively regular, corresponding to quiescent disks with sufficiently low accretion flow, allowing for existence of nearly uniform magnetic field near the horizon. This really corresponds to the so called hard states where the HF QPOs are observed \cite{Fen-Bel-Gal:2004:MNRAS:}. In the so called soft states, related to high accretion rates and relatively irregular disks, no HF QPOs are observed, thus we can expect that the uniformity of the magnetic field in the horizon vicinity becomes strongly violated by the strong irregular accretion flow.
\end{itemize}

The charged test particle oscillatory frequencies are sensitive to any magnetic field presence. The applicability of the magnetized HF QPO ER and RP models for QPOs observed in completely different sources allows us to conclude that the proposed models with inclusion of the effect of magnetic field can be considered as one of the possible explanations of the HF QPOs occurring in the field of microquasars. Assuming the main source of synchrotron radiation producing X-rays are the relativistic electrons, we estimate the magnetic field in the vicinity of the black hole in the three sources to be of order $10^{-5}$~G which can serve as possible signature of the Galactic magnetic field magnitude. For heavier particles (ions, charged hotspots) much larger magnetic fields are necessary for fitting the data. The model also gives limits on character of oscillating matter and intenzity of the magnetic field.

\section*{Acknowledgments}

The authors acknowledge the Silesian University in Opava Grant No. SGS/14/2016. M.K. and A.T. acknowledge the Czech Science Foundation Grant No. 16-03564Y, Z.S. acknowledges the Albert Einstein Centre for Gravitation and Astrophysics supported by the Czech Science Foundation Grant No. 14-37086G. 



\def\prc{Phys. Rev. C}
\def\pre{Phys. Rev. E}
\def\prd{Phys. Rev. D}
\def\jcap{Journal of Cosmology and Astroparticle Physics}
\def\apss{Astrophysics and Space Science}
\def\mnras{Monthly Notices of the Royal Astronomical Society}
\def\apj{The Astrophysical Journal}
\def\aap{Astronomy and Astrophysics}
\def\actaa{Acta Astronomica}
\def\pasj{Publications of the Astronomical Society of Japan}
\def\apjl{Astrophysical Journal Letters}
\def\pasa{Publications Astronomical Society of Australia}
\def\nat{Nature}
\def\physrep{Physics Reports}
\def\araa{Annual Review of Astronomy and Astrophysics}
\def\apjs{The Astrophysical Journal Supplement}
\def\aapr{The Astronomy and Astrophysics Review}
\def\procspie{Proceedings of the SPIE}


\begin{thebibliography}{10}

\bibitem{Abr-Klu:2001:AA:}
M.~A. {Abramowicz} and W.~{Klu{\'z}niak}.
\newblock {A precise determination of black hole spin in GRO J1655-40}.
\newblock {\em \aap}, 374:L19--L20, August 2001.

\bibitem{Ali-Gal:1981:GRG:}
A.~N. {Aliev} and D.~V. {Galtsov}.
\newblock {Radiation from relativistic particles in nongeodesic motion in a
  strong gravitational field}.
\newblock {\em General Relativity and Gravitation}, 13:899--912, October 1981.

\bibitem{Bak-etal:2012:CLAQG:}
P.~{Bakala}, M.~{Urbanec}, E.~{{\v S}r{\'a}mkov{\'a}}, Z.~{Stuchl{\'{\i}}k},
  and G.~{T{\"o}r{\"o}k}.
\newblock {On magnetic-field-induced corrections to the orbital and epicyclic
  frequencies: paper II. Slowly rotating magnetized neutron stars}.
\newblock {\em Classical and Quantum Gravity}, 29(6):065012, March 2012.

\bibitem{Bak-etal:2010:CLAQG:}
P.~{Bakala}, E.~{{\v S}r{\'a}mkov{\'a}}, Z.~{Stuchl{\'{\i}}k}, and
  G.~{T{\"o}r{\"o}k}.
\newblock {On magnetic-field-induced non-geodesic corrections to relativistic
  orbital and epicyclic frequencies}.
\newblock {\em Classical and Quantum Gravity}, 27(4):045001, February 2010.

\bibitem{Bal-Bic-Stu:1989:BAC:}
V.~{Balek}, J.~{Bi{\v{c}}{\'a}k}, and Z.~{Stuchl{\'{\i}}k}.
\newblock {The motion of the charged particles in the field of rotating charged
  black holes and naked singularities. II - The motion in the equatorial
  plane}.
\newblock {\em Bulletin of the Astronomical Institutes of Czechoslovakia},
  40:133--165, June 1989.

\bibitem{Bar-Oli-Mil:2005:MONNR:}
D.~{Barret}, J.-F. {Olive}, and M.~C. {Miller}.
\newblock {An abrupt drop in the coherence of the lower kHz quasi-periodic
  oscillations in 4U 1636-536}.
\newblock {\em \mnras}, 361:855--860, August 2005.

\bibitem{Bec-Wie:2013:book}
R.~{Beck} and R.~{Wielebinski}.
\newblock {\em {Magnetic Fields in Galaxies}}, page 641.
\newblock 2013.

\bibitem{Bel-Alt:2013:MNRAS}
T.~M. {Belloni} and D.~{Altamirano}.
\newblock {High-frequency quasi-periodic oscillations from GRS 1915+105}.
\newblock {\em \mnras}, 432:10--18, June 2013.

\bibitem{Bell-etal:2012:MNRAS}
T.~M. {Belloni}, A.~{Sanna}, and M.~{M{\'e}ndez}.
\newblock {High-frequency quasi-periodic oscillations in black hole binaries}.
\newblock {\em \mnras}, 426:1701--1709, November 2012.

\bibitem{Bic-Stu-Bal:1989:BAC:}
J.~{Bi{\v{c}}{\'a}k}, Z.~{Stuchl{\'{\i}}k}, and V.~{Balek}.
\newblock {The motion of charged particles in the field of rotating charged
  black holes and naked singularities}.
\newblock {\em Bulletin of the Astronomical Institutes of Czechoslovakia},
  40:65--92, March 1989.

\bibitem{Bur:2011:POS:}
M.~{Bursa}.
\newblock {Epicyclic Frequencies, Resonances {\&} QPOs}.
\newblock In {\em Fast X-ray Timing and Spectroscopy at Extreme Count Rates
  (HTRS 2011)}, page~33, 2011.

\bibitem{Cre-etal:2013:ApJS:}
C.~{Cremaschini}, J.~{Kov{\'a}{\v r}}, P.~{Slan{\'y}}, Z.~{Stuchl{\'{\i}}k},
  and V.~{Karas}.
\newblock {Kinetic Theory of Equilibrium Axisymmetric Collisionless Plasmas in
  Off-equatorial Tori around Compact Objects}.
\newblock {\em \apjs}, 209:15, November 2013.

\bibitem{Eat-etal:2013:NATUR:}
R.~P. {Eatough} and et~al.
\newblock {A strong magnetic field around the supermassive black hole at the
  centre of the Galaxy}.
\newblock {\em \nat}, 501:391--394, September 2013.

\bibitem{Fen-Bel-Gal:2004:MNRAS:}
R.~P. {Fender}, T.~M. {Belloni}, and E.~{Gallo}.
\newblock {Towards a unified model for black hole X-ray binary jets}.
\newblock {\em \mnras}, 355:1105--1118, December 2004.

\bibitem{Fender-etal:1997:MNRAS:}
R.~P. {Fender}, G.~G. {Pooley}, C.~{Brocksopp}, and S.~J. {Newell}.
\newblock {Rapid infrared flares in GRS 1915+105: evidence for infrared
  synchrotron emission}.
\newblock {\em \mnras}, 290:L65--L69, October 1997.

\bibitem{Fro-Sho:2010:PHYSR4:}
V.~P. {Frolov} and A.~A. {Shoom}.
\newblock {Motion of charged particles near a weakly magnetized Schwarzschild
  black hole}.
\newblock {\em \prd}, 82(8):084034, October 2010.

\bibitem{Fro-Sho-Tzo:2014:PRD:}
V.~P. {Frolov}, A.~A. {Shoom}, and C.~{Tzounis}.
\newblock {Radiation from an emitter revolving around a magnetized nonrotating
  black hole}.
\newblock {\em \prd}, 90(2):024027, July 2014.

\bibitem{Fro-Sho-Tzo:2014:JCAP:}
V.~P. {Frolov}, A.~A. {Shoom}, and C.~{Tzounis}.
\newblock {Spectral line broadening in magnetized black holes}.
\newblock {\em \jcap}, 7:059, July 2014.

\bibitem{Fu-Lai:2009:ApJ:}
W.~{Fu} and D.~{Lai}.
\newblock {Effects of Magnetic Fields on the Diskoseismic Modes of Accreting
  Black Holes}.
\newblock {\em \apj}, 690:1386--1392, January 2009.

\bibitem{Gre-Pre-Poh:1995:AAP:}
J.~{Greiner}, P.~{Predehl}, and M.~{Pohl}.
\newblock {ROSAT observations of GRO J1655-40}.
\newblock {\em \aap}, 297:L67, May 1995.

\bibitem{Hao-Zha:2009:APJ:}
J.~F. {Hao} and S.~N. {Zhang}.
\newblock {Large-scale Cavities Surrounding Microquasars Inferred from
  Evolution of Their Relativistic Jets}.
\newblock {\em \apj}, 702:1648--1661, September 2009.

\bibitem{Kol-Stu-Tur:2015:CLAQG:}
M.~{Kolo{\v s}}, Z.~{Stuchl{\'{\i}}k}, and A.~{Tursunov}.
\newblock {Quasi-harmonic oscillatory motion of charged particles around a
  Schwarzschild black hole immersed in a uniform magnetic field}.
\newblock {\em Classical and Quantum Gravity}, 32(16):165009, August 2015.

\bibitem{Kon-Liu:2012:PRD:}
R.~A. {Konoplya} and Y.-C. {Liu}.
\newblock {Motion of charged particles and quasinormal modes around the
  magnetically and tidally deformed black hole}.
\newblock {\em \prd}, 86(8):084007, October 2012.

\bibitem{Kop-Kar:2014:APJ:}
O.~{Kop{\'a}{\v c}ek} and V.~{Karas}.
\newblock {Inducing Chaos by Breaking Axial Symmetry in a Black Hole
  Magnetosphere}.
\newblock {\em \apj}, 787:117, June 2014.

\bibitem{Kos-etal:2009:ASTRA:}
U.~{Kosti{\'c}}, A.~{{\v C}ade{\v z}}, M.~{Calvani}, and A.~{Gomboc}.
\newblock {Tidal effects on small bodies by massive black holes}.
\newblock {\em \aap}, 496:307--315, March 2009.

\bibitem{Kov-etal:2010:CLAQG:}
J.~{Kov{\'a}{\v r}}, O.~{Kop{\'a}{\v c}ek}, V.~{Karas}, and
  Z.~{Stuchl{\'{\i}}k}.
\newblock {Off-equatorial orbits in strong gravitational fields near compact
  objects - II: halo motion around magnetic compact stars and magnetized black
  holes}.
\newblock {\em Classical and Quantum Gravity}, 27(13):135006, July 2010.

\bibitem{McC-etal:2011:CLAQG:}
J.~E. {McClintock}, R.~{Narayan}, S.~W. {Davis}, L.~{Gou}, A.~{Kulkarni}, J.~A.
  {Orosz}, R.~F. {Penna}, R.~A. {Remillard}, and J.~F. {Steiner}.
\newblock {Measuring the spins of accreting black holes}.
\newblock {\em Classical and Quantum Gravity}, 28(11):114009, June 2011.

\bibitem{McCli-Rem:2004:CompactX-Sources:}
J.~E. {McClintock} and R.~A. {Remillard}.
\newblock {\em {Black hole binaries}}, pages 157--213.
\newblock April 2006.

\bibitem{Mon-Zan:2012:MNRAS:}
P.~J. {Montero} and O.~{Zanotti}.
\newblock {Oscillations of relativistic axisymmetric tori and implications for
  modelling kHz-QPOs in neutron star X-ray binaries}.
\newblock {\em \mnras}, 419:1507--1514, January 2012.

\bibitem{Mot-etal:2014a:MNRAS:}
S.~E. {Motta}, T.~M. {Belloni}, L.~{Stella}, T.~{Mu{\~n}oz-Darias}, and
  R.~{Fender}.
\newblock {Precise mass and spin measurements for a stellar-mass black hole
  through X-ray timing: the case of GRO J1655-40}.
\newblock {\em \mnras}, 437:2554--2565, January 2014.

\bibitem{Pug-Que-Ruf:2011:PHYSR4:}
D.~{Pugliese}, H.~{Quevedo}, and R.~{Ruffini}.
\newblock {Equatorial circular motion in Kerr spacetime}.
\newblock {\em \prd}, 84(4):044030, August 2011.

\bibitem{Pug-Que-Ruf:2017:EPJC:}
D.~{Pugliese}, H.~{Quevedo}, and R.~{Ruffini}.
\newblock {General classification of charged test particle circular orbits in
  Reissner-Nordstr{\"o}m spacetime}.
\newblock {\em European Physical Journal C}, 77:206, April 2017.

\bibitem{Rem:2005:ASTRN:}
R.~A. {Remillard}.
\newblock {X-ray spectral states and high-frequency QPOs in black hole
  binaries}.
\newblock {\em Astronomische Nachrichten}, 326:804--807, November 2005.

\bibitem{Rem-McCli:2006:ARAA:}
R.~A. {Remillard} and J.~E. {McClintock}.
\newblock {X-Ray Properties of Black-Hole Binaries}.
\newblock {\em \araa}, 44:49--92, September 2006.

\bibitem{Rez-etal:2003:MNRAS:}
L.~{Rezzolla}, S.~{Yoshida}, T.~J. {Maccarone}, and O.~{Zanotti}.
\newblock {A new simple model for high-frequency quasi-periodic oscillations in
  black hole candidates}.
\newblock {\em \mnras}, 344:L37--L41, September 2003.

\bibitem{Sha-etal:2006:ApJ:}
R.~{Shafee}, J.~E. {McClintock}, R.~{Narayan}, S.~W. {Davis}, L.-X. {Li}, and
  R.~A. {Remillard}.
\newblock {Estimating the Spin of Stellar-Mass Black Holes by Spectral Fitting
  of the X-Ray Continuum}.
\newblock {\em \apjl}, 636:L113--L116, January 2006.

\bibitem{Ste:2014:MNRAS:}
I.~Z. {Stefanov}.
\newblock {Confronting models for the high-frequency QPOs with Lense-Thirring
  precession}.
\newblock {\em \mnras}, 444:2178--2185, November 2014.

\bibitem{Ste-Vie:1998:ApJ:}
L.~{Stella} and M.~{Vietri}.
\newblock {Lense-Thirring Precession and Quasi-periodic Oscillations in
  Low-Mass X-Ray Binaries}.
\newblock {\em \apjl}, 492:L59, January 1998.

\bibitem{Ste-Vie:1999:PHYSRL:}
L.~{Stella} and M.~{Vietri}.
\newblock {kHz Quasiperiodic Oscillations in Low-Mass X-Ray Binaries as Probes
  of General Relativity in the Strong-Field Regime}.
\newblock {\em Physical Review Letters}, 82:17--20, January 1999.

\bibitem{Ste-Vie-Mor:1999:ApJ:}
L.~{Stella}, M.~{Vietri}, and S.~M. {Morsink}.
\newblock {Correlations in the Quasi-periodic Oscillation Frequencies of
  Low-Mass X-Ray Binaries and the Relativistic Precession Model}.
\newblock {\em \apjl}, 524:L63--L66, October 1999.

\bibitem{Stu-Bic-Bal:1999:GRG:}
Z.~{Stuchl{\'{\i}}k}, J.~{Bi{\v{c}}{\'a}k}, and V.~{Balek}.
\newblock {The Shell of Incoherent Charged Matter Falling onto a Charged
  Rotating Black Hole}.
\newblock {\em General Relativity and Gravitation}, 31:53--71, January 1999.

\bibitem{Stu-Kol:2012:JCAP:}
Z.~{Stuchl{\'{\i}}k} and M.~{Kolo{\v s}}.
\newblock {String loops in the field of braneworld spherically symmetric black
  holes and naked singularities}.
\newblock {\em \jcap}, 10:008, October 2012.

\bibitem{Stu-Kol:2014:PHYSR4:}
Z.~{Stuchl{\'{\i}}k} and M.~{Kolo{\v s}}.
\newblock {String loops oscillating in the field of Kerr black holes as a
  possible explanation of twin high-frequency quasiperiodic oscillations
  observed in microquasars}.
\newblock {\em \prd}, 89(6):065007, March 2014.

\bibitem{Stu-Kol:2015:MNRAS:}
Z.~{Stuchl{\'{\i}}k} and M.~{Kolo{\v s}}.
\newblock {Mass of intermediate black hole in the source M82 X-1 restricted by
  models of twin high-frequency quasi-periodic oscillations}.
\newblock {\em \mnras}, 451:2575--2588, August 2015.

\bibitem{Stu-Kol:2016:EPJC:}
Z.~{Stuchl{\'{\i}}k} and M.~{Kolo{\v s}}.
\newblock {Acceleration of the charged particles due to chaotic scattering in
  the combined black hole gravitational field and asymptotically uniform
  magnetic field}.
\newblock {\em European Physical Journal C}, 76:32, January 2016.

\bibitem{Stu-Kol:2016:APJ:}
Z.~{Stuchl{\'{\i}}k} and M.~{Kolo{\v s}}.
\newblock {Controversy of the GRO J1655-40 Black Hole Mass and Spin Estimates
  and Its Possible Solutions}.
\newblock {\em \apj}, 825:13, July 2016.

\bibitem{Stu-Kol:2016:ASTRA:}
Z.~{Stuchl{\'{\i}}k} and M.~{Kolo{\v s}}.
\newblock {Models of quasi-periodic oscillations related to mass and spin of
  the GRO J1655-40 black hole}.
\newblock {\em \aap}, 586:A130, February 2016.

\bibitem{Stu-Kot-Tor:2011:ASTRA:}
Z.~{Stuchl{\'{\i}}k}, A.~{Kotrlov{\'a}}, and G.~{T{\"o}r{\"o}k}.
\newblock {Resonant radii of kHz quasi-periodic oscillations in Keplerian discs
  orbiting neutron stars}.
\newblock {\em \aap}, 525:A82, January 2011.

\bibitem{Stu-Kot-Tor:2013:ASTRA:}
Z.~{Stuchl{\'{\i}}k}, A.~{Kotrlov{\'a}}, and G.~{T{\"o}r{\"o}k}.
\newblock {Multi-resonance orbital model of high-frequency quasi-periodic
  oscillations: possible high-precision determination of black hole and neutron
  star spin}.
\newblock {\em \aap}, 552:A10, April 2013.

\bibitem{Stu-etal:2015:ACTA:}
Z.~{Stuchl{\`i}k}, M.~{Urbanec}, A.~{Kotrlov{\`a}}, G.~{T{\"o}r{\"o}k}, and
  K.~{Goluchov{\`a}}.
\newblock {Equations of State in the Hartle-Thorne Model of Neutron Stars
  Selecting Acceptable Variants of the Resonant Switch Model of Twin HF QPOs in
  the Atoll Source 4U 1636-53}.
\newblock {\em \actaa}, 65:169--195, June 2015.

\bibitem{Tor-etal:2005:ASTRA:}
G.~{T{\"o}r{\"o}k}, M.~A. {Abramowicz}, W.~{Klu{\'z}niak}, and
  Z.~{Stuchl{\'{\i}}k}.
\newblock {The orbital resonance model for twin peak kHz quasi periodic
  oscillations in microquasars}.
\newblock {\em \aap}, 436:1--8, June 2005.

\bibitem{Tor-etal:2011:ASTRA:}
G.~{T{\"o}r{\"o}k}, A.~{Kotrlov{\'a}}, E.~{{\v S}r{\'a}mkov{\'a}}, and
  Z.~{Stuchl{\'{\i}}k}.
\newblock {Confronting the models of 3:2 quasiperiodic oscillations with the
  rapid spin of the microquasar GRS 1915+105}.
\newblock {\em \aap}, 531:A59, July 2011.

\bibitem{Tur-Stu-Kol:2016:PRD:}
A.~{Tursunov}, Z.~{Stuchl{\'{\i}}k}, and M.~{Kolo{\v s}}.
\newblock {Circular orbits and related quasiharmonic oscillatory motion of
  charged particles around weakly magnetized rotating black holes}.
\newblock {\em \prd}, 93(8):084012, April 2016.

\bibitem{Wald:1974:PHYSR4:}
R.~M. {Wald}.
\newblock {Black hole in a uniform magnetic field}.
\newblock {\em \prd}, 10:1680--1685, September 1974.

\bibitem{Wald:1984:book:}
R.~M. {Wald}.
\newblock {\em {General relativity}}.
\newblock University of Chicago Press, Chicago, 1984.

\bibitem{Wan-Dai-Lu:2005:APSS:}
X.~Y. {Wang}, Z.~G. {Dai}, and T.~{Lu}.
\newblock {The Large-Scale, Decelerating X-ray Jets from the Microquasar Xte
  J1550-564: Evidence for External Shocks Caused by the Jet-Ism Interaction?}
\newblock {\em \apss}, 297:155--166, June 2005.

\bibitem{Zak-etal:2003:MNRAS}
A.~F. {Zakharov}, N.~S. {Kardashev}, V.~N. {Lukash}, and S.~V. {Repin}.
\newblock {Magnetic fields in active galactic nuclei and microquasars}.
\newblock {\em \mnras}, 342:1325--1333, July 2003.

\end{thebibliography}

\end{document}